\documentclass[acmtog, nonacm]{acmart}
\acmSubmissionID{109}

\usepackage{booktabs} 
\usepackage{xcolor}

\usepackage{mathtools}

\newcommand\todommc[1]{{\color{red}{#1}}}
\usepackage{xcolor}

\usepackage[dvipsnames]{xcolor}
\usepackage{amsmath}
\usepackage{algpseudocode}
\usepackage{tabularx}
\usepackage{bm}
\usepackage{xcolor}
\usepackage{cancel}
\usepackage{calc}
\usepackage{accents}
\usepackage{tikz}
\usepackage{pifont}
\usetikzlibrary{backgrounds}
\usetikzlibrary{decorations.pathreplacing, arrows.meta, tikzmark}
\usetikzlibrary{calc}
\usetikzlibrary {shapes.geometric} 
\usepackage{wrapfig}
\usetikzlibrary {fit}
\usepackage{amsmath}
\usepackage{cleveref}
\usepackage{tikzscale}
\usepackage{graphicx}
\usepackage{units} 
\usepackage[percent]{overpic}
\usepackage{subfig}
\usepackage[export]{adjustbox}
\usepackage{rotating} 

\usepackage{pgfplots}
\usepgfplotslibrary{groupplots}
\pgfplotsset{compat=1.18}

\usepackage[table,xcdraw]{xcolor}
\usepackage{booktabs} 
\usepackage{makecell}

\tikzset{every node/.style={font=\fontfamily{LinuxBiolinumT-TLF}\selectfont}}
\usepackage{bm}
\usetikzlibrary{spy}

\newcommand{\revision}[1]{\textcolor{black}{#1}}

\usepackage[normalem]{ulem}
\usepackage{xcolor}

\usepackage{cancel}

\newcommand{\revisionstrike}[1]{%
  \ifmmode
    \textcolor{red}{\cancel{#1}}%
  \else
    \textcolor{red}{\sout{#1}}%
  \fi
}

\newlength{\subcolumnwidth}

\newcommand{\nextsubcolumn}[1][]{%
  \cr\noalign{\hfill}
  \if\relax\detokenize{#1}\relax\else\hsize=#1\setlength{\subcolumnwidth}{\hsize}\fi
}

\usepackage[ruled]{algorithm2e} 

\SetAlFnt{\small}
\SetAlCapFnt{\small}
\SetAlCapNameFnt{\small}
\SetAlCapHSkip{0pt}

\usepackage{tikzducks}
\usepackage{duckuments}

\makeatletter
\newcommand{\showfontinfo}{
    Font: \f@family \f@series \f@shape, Size: \f@size
}
\makeatother
\begin{document}
\title{Low-Rank Koopman Deformables with Log-Linear Time Integration}

\author{Yue Chang}
\orcid{0000-0002-2587-827X}
\affiliation{%
  \institution{University of Toronto}
  \country{Canada}}
\email{changyue.chang@mail.utoronto.ca}

\author{Peter Yichen Chen}
\orcid{0000-0003-1919-5437}
\affiliation{%
  \institution{University of British Columbia}
  \country{Canada}}
\email{peter.chen@ubc.ca}

\author{Eitan Grinspun}
\orcid{0000-0003-4460-7747}
\affiliation{%
  \institution{University of Toronto}
  \country{Canada}}
\email{eitan@cs.toronto.edu}

\author{Maurizio M. Chiaramonte}
\orcid{0000-0002-2529-3159}
\affiliation{%
  \institution{Meta Reality Labs Research}
  \country{USA}}
\email{mchiaram@meta.com}

\begin{abstract}
We present a low-rank Koopman operator formulation for accelerating deformable subspace simulation. Using a Dynamic Mode Decomposition (DMD) parameterization of the Koopman operator, our method learns the temporal evolution of deformable dynamics and predicts future states through efficient matrix evaluations instead of sequential time integration. This yields log-linear scaling in the number of time steps and allows large portions of the trajectory to be skipped while retaining accuracy. The resulting temporal efficiency is especially advantageous for optimization tasks such as control and initial-state estimation, where the objective often depends largely on the final configuration.

To broaden the scope of Koopman-based reduced-order models in graphics, we introduce a discretization-agnostic extension that learns shared dynamic behavior across multiple shapes and mesh resolutions. Prior DMD-based approaches have been restricted to a single shape and discretization, which limits their usefulness for tasks involving geometry variation. Our formulation generalizes across both shape and discretization, which enables fast shape optimization that was previously impractical for DMD models. This expanded capability highlights the potential of Koopman operator learning as a practical tool for efficient deformable simulation and design.
\end{abstract}

%
%


%
%


\begin{teaserfigure}
    \centering
    \includegraphics[width=1.0\linewidth]{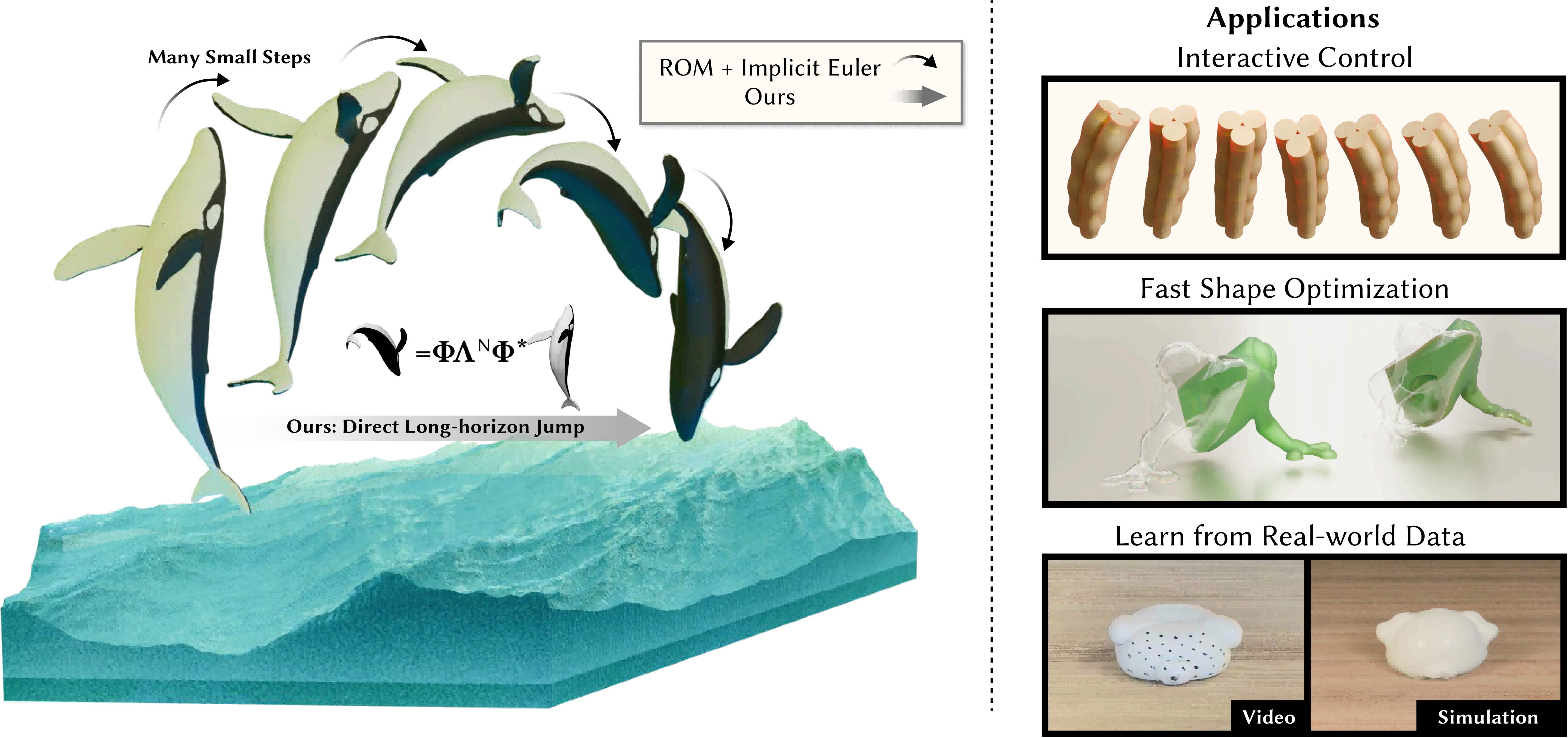}
    \caption{We present a Koopman-based reduced model for deformable dynamics, enabling direct long-horizon integration via matrix exponentiation. 
Unlike traditional reduced models that require many small steps and costly Newton solves (e.g., implicit Euler), our formulation enables a single-step jump over long time horizons while preserving accurate deformation trajectories. 
This efficient integration unlocks a range of downstream applications (right), including real-time interactive control, shape optimization, and learning from real-world video data without requiring access to a simulator.}
    \label{fig:teaser}
\end{teaserfigure}

\maketitle
\newif\ifmaurizio
\mauriziofalse

\begin{figure*}
    \centering
    \includegraphics[width=1.0\linewidth]{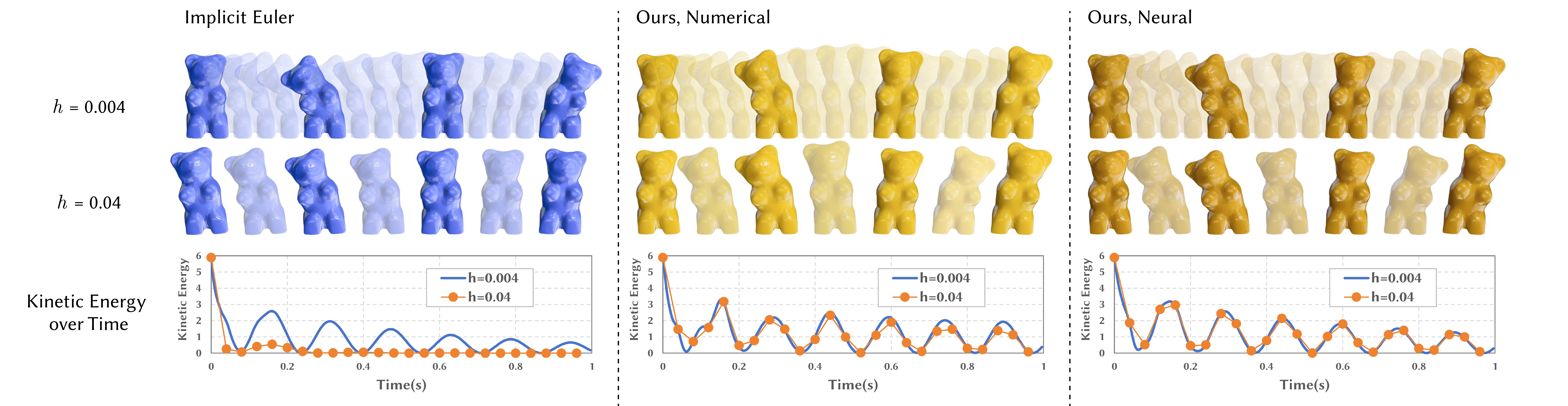}
    \caption{\emph{Behavior under larger time steps.}
We compare deformation and kinetic energy for implicit Euler and our method (both numerical and neural) at two time step sizes.
Because our dynamics are represented by a linear operator, increasing the time step corresponds to exponentiating this operator, yielding nearly identical behavior across step sizes (middle and right, yellow).
This computation scales only logarithmically with the step size.
In contrast, linear model reduction integrated with implicit Euler, requiring Newton solves at each step, exhibits significantly different and strongly damped behavior as the time step increases (left).
}
    \label{fig:gummybear_comparison}
\end{figure*}

\section{Introduction}
\label{sec:introduction}

Deformable objects are ubiquitous in the physical world, and their real-time simulation is central to applications ranging from robotics and control to AR/VR and interactive design. 
Among existing approaches, reduced-order modeling has become one of the most effective tools for achieving real-time performance, from early linear-algebra–based techniques~\cite{barbivc2005real,An:Cubature:2008} to more recent neural formulations~\cite{Modi:2024:Simplicits,fulton2019latent,lyu2024accelerate}. 
These methods accelerate simulation by projecting high-dimensional deformable dynamics onto a lower-dimensional manifold that captures the dominant spatial degrees of freedom.

However, deformable dynamics are not only high-dimensional in space, but also evolve over time. 
While most reduced-order methods focus on spatial compression, time integration is still performed step by step, and the computational cost scales linearly with the number of time steps. 
In this work, we address this limitation by reducing not only the spatial complexity of deformable simulation, but also its temporal complexity through efficient time stepping in reduced dynamical systems.

We calculate a low-rank linear approximation of the time integrator, which can be interpreted as learning a Koopman operator for deformable dynamics. 
We adopt dynamic mode decomposition (DMD) ~\cite{schmid2010dmd, Kutz2015mrDMD, sashidhar2022bagging} to approximate time integration in reduced deformable systems.

Applying Koopman-style linearization to deformable systems presents challenges that differ from prior reduced-order models. 
While most reduced-order models (ROMs) for deformable simulation reduce only spatial displacement~\cite{Xu:2016:PSS,shen2021high}, and existing DMD formulations for fluids operate solely on velocity fields~\cite{chen2025dmd}, deformable dynamics require the coupled evolution of both displacement and velocity to faithfully capture elastic motion. 
We therefore construct a joint state representation that enables accurate linearized time stepping of deformable dynamics.

We extend the Koopman operator to shape space, enabling consistent modeling across a family of deformable geometries. We proposed a dynamical model that generalizes across both shapes and discretizations, implemented via neural representation. This shape-aware formulation allows a single model to operate across meshes of varying resolution, enabling applications such as fast shape optimization that are infeasible with purely numerical DMD methods. Recent discretization-agnostic approaches in graphics~\cite{chang:2023:licrom,Modi:2024:Simplicits,chang2024neuralrepresentationshapedependentlaplacian} demonstrate resolution-independent behavior in spatial operators; our formulation complements these efforts by additionally supporting accelerated time integration through joint spatial–temporal reduction.

A crucial benefit of this formulation is that time stepping reduces to exponentiating a small matrix, yielding log-linear complexity with respect to the number of time steps. 
This enables fast integration even over long time horizons, making the method suitable for interactive simulation as well as control and optimization tasks. 
We further show how this efficient time stepping supports inverse problems such as control-force estimation and shape optimization, which can be performed quickly or even interactively using our framework.

\paragraph{Contributions.}
In summary, we:
\begin{itemize}
    \item introduce a DMD (Koopman) formulation for deformable simulation in computer graphics;
    \item propose a discretization-agnostic neural extension that enables generalization across different meshes and shapes;
    \item demonstrate the benefits of log-linear time stepping through real-time interaction, interactive control-force estimation, and fast shape optimization.
\end{itemize}

\section{Related Work}
\subsection{Reduced-order Simulation}
Reduced-order methods have been widely used to accelerate physical simulation across many domains, including fluids~\cite{Cui2018Eigenfluids,deWitt2012Eigenfunctions}, sound~\cite{james2006precomputed}, and collision handling~\cite{Barbic2010SubspaceSelfCollision}. 

Our work is most closely related to reduced-order methods for deformable simulation~\cite{barbivc2005real,Mukherjee:2016:IDSR,trusty:2023:subspacemfem,Tycowicz:ECRDO:2013,Kim:Skipping:2009}. 
Among them, Barbič and James~\cite{barbivc2005real} proposed a polynomial force approximation that captures the physics of the system in addition to reducing the spatial degrees of freedom. 
Similarly, we approximate the deformable dynamics, but unlike polynomial force models, our formulation operates on a joint state of displacement and velocity and represents the time evolution through a linear operator rather than a nonlinear force law.
Kim et al.~\cite{Kim:Skipping:2009} proposed a step-skipping method that adaptively reduces the spatial degrees of freedom from full space to reduced space, which is a different notion of skipping than ours. 
Their method enables efficient per-step computation by constructing bases on the fly, avoiding offline precomputation, but it still relies on numerical time integration to advance the reduced system, so the overall cost scales linearly with the number of time steps.

A key distinction of our approach is that time evolution itself is reduced. 
By learning a Koopman-based linear time-propagation operator in the reduced space, we decouple temporal evolution from numerical integration and advance the system by exponentiating a low-rank matrix, enabling time stepping that scales logarithmically with the simulated time horizon. 
This temporal reduction is particularly important for interactive control and optimization, where fast evaluation of long time intervals is required.

\subsection{Neural Physics Simulation}

Neural networks have recently become an important tool for accelerating and approximating physical simulation. 
They have been applied to a broad range of problems, including cloth and soft-body simulation~\cite{Bertiche:2022:NCS,kair2024neuralclothsim,Zhang:2024:NGDSR, chenwu:2023:insr-pde}, fluid simulation~\cite{kim2019deep,Chu2022Physics,deng2023neural,Wang2024PICT,10.1145/3641519.3657438,tao2024neural}, and collision handling~\cite{Romero:LCCHSD:2021, romero2023contactdescriptorlearning, yang2020learning,Cai:2022:CSDF, zesch2023neural}. 

A related line of work applies neural networks to reduced-order deformable simulation. 
For example, Fulton et al.~\cite{fulton2019latent} and subsequent extensions~\cite{shen2021high,Sharp:2023:datafree,lyu2024accelerate} introduce learned nonlinear bases to better capture complex deformation. 
These approaches aim to improve expressiveness within reduced coordinates, but they still rely on conventional time integration and are typically trained for a fixed mesh or shape. 
In contrast, our method preserves a physically meaningful state representation and focuses on learning a linear time-evolution operator, enabling efficient long-horizon prediction and generalization across discretizations. 
Both~\cite{Holden:2019:SNP} and~\cite{Li2025LatentDynamics} learn the physics of the reduced system directly, rather than relying on conventional time integration. 
However, their learned physics is represented by nonlinear operators, whereas ours is represented by a linear operator that admits efficient exponentiation.

Neural fields have further enabled discretization-agnostic representations for deformable simulation by modeling spatial quantities as continuous functions of position~\cite{chen2023crom,chang:2023:licrom,Modi:2024:Simplicits,chang2024neuralrepresentationshapedependentlaplacian}. 
However, their time stepping is still performed using conventional numerical integrators. 
Our discretization-agnostic neural Koopman formulation additionally enables fast time stepping while preserving the discretization-agnostic property.

\subsection{Dynamic Mode Decomposition}
Our linear approximation of the time-evolution operator is implemented using Dynamic Mode Decomposition (DMD). 
Originally introduced by Schmid~\cite{schmid2010dynamic}, DMD has since been extended in a variety of directions, including variable projection~\cite{askham2018variable}, ensemble and bagging methods~\cite{sashidhar2022bagging}, multiresolution formulations~\cite{Kutz2015mrDMD}, as well as many others~\cite{mezic2020koopman}.
Our numerical implementation of deformable simulation builds on the \texttt{PyDMD} library~\cite{Ichinaga2024}, which provides a reference implementation of the classical DMD formulation~\cite{schmid2010dynamic}.

Within the graphics community, DMD was recently introduced for fluid simulation by Chen et al.~\cite{chen2025dmd}, where it is used to model the temporal advection of velocity fields and enables faster-than-linear time integration. 
Our work targets a different domain, deformable simulation, and extends the applicability of DMD to elastic dynamics, complementing this line of research.

Although formulated differently, de Aguiar et al.~\cite{deAguiar2010Stable} also represent cloth dynamics using a linear approximation conditioned on character motion. 
Similarly, we represent deformable physics through a linearized model. 
However, their approach is pose-conditioned, which introduces nonlinear dynamics during motion and precludes time stepping via operator exponentiation. 
In addition, their formulation is tied to a single training mesh, whereas our discretization-agnostic extension enables consistent behavior across different discretizations.

\begin{figure}
\centering
\includegraphics[width = 1.0\linewidth]{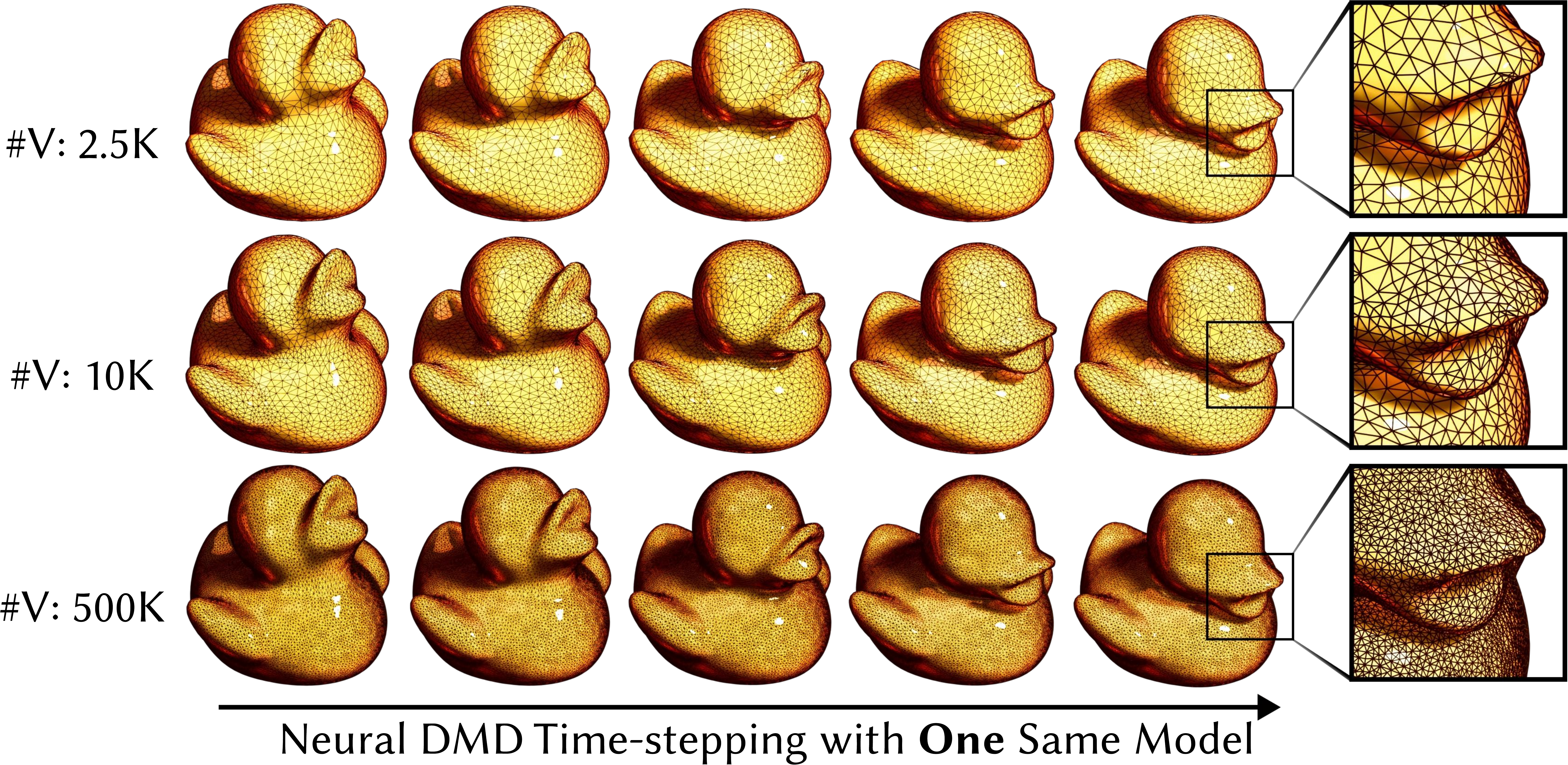}
\caption{\emph{Generalization across discretizations.}
Our neural formulation provides a discretization-agnostic extension of the numerical DMD model, allowing a single model to represent the dynamics of meshes with widely varying resolutions (2.5k, 10k, and 500k vertices).}
\label{fig:duck_discretization_agnostic}
\end{figure}
\newif\ifmaurizio
\mauriziotrue
\ifmaurizio
\section{Background: Low-Rank Approximation of the Koopman Operator}

\def\Kc{\mathcal{K}}
\def\Vc{\mathcal{V}}
\def\Mc{\mathcal{M}}
\def\Rbb{\mathbb{R}}
\def\x{\boldsymbol{x}}
\def\Fb{\boldsymbol{F}}
\def\ub{\boldsymbol{u}}

Let the state space of a dynamical system is given by $\Mc$, where $\ub \in \Mc$ denotes a specific state of the system
(for example the positions and velocities of an elastic body). Consider a discrete-time dynamical system
\begin{equation}
\ub_{n+1} = \Fb(\ub_n).
\end{equation}
Rather than studying the nonlinear evolution of the state directly, Koopman operator theory studies the evolution of
\emph{observables}, i.e., scalar-valued functions $g \in \Vc$ where
\[
\Vc := \{ g \mid g : \Mc \to \Rbb \}.
\]
The Koopman operator $\Kc : \Vc \to \Vc$ is defined by composition with the dynamics,
\begin{equation}
(\Kc g)(\ub_n) = g(\Fb(\ub_n)) = g(\ub_{n+1}).
\end{equation}
\revision{Importantly, $\Kc$ is linear and infinite dimensional even when $\Fb$ is nonlinear and finite dimensional.}

If we define the Koopman eigenfunctions and eigenvalues by $\varphi_i \in \Vc$ and $\lambda_i \in \mathbb{C}$, respectively, then
\begin{equation}
\Kc \varphi_i(\ub)  = \lambda_i \varphi_i(\ub), \quad i = 1,2,\ldots
\end{equation}
and, at least formally, any observable $g$ can be expanded in the Koopman eigenfunction basis, for $\alpha_i \in \mathbb{C}$, as
\begin{equation}\label{eq:koopman_expansion}
g(\ub) = \sum_{i=1}^\infty \alpha_i \varphi_i(\ub).
\end{equation}
Applying $\Kc$ to \eqref{eq:koopman_expansion} yields
\begin{equation}
(\Kc g)(\ub_k) = g(\ub_{k+1}) = \sum_{i=1}^\infty \alpha_i \lambda_i \varphi_i(\ub_k),
\end{equation}
which shows that, in Koopman eigenfunction coordinates, each component evolves independently by multiplication with its eigenvalue.
A natural way to obtain a finite-dimensional approximation would be to truncate this expansion to $m$ terms, i.e.,
\[
g(\ub) \approx \sum_{i=1}^m \alpha_i \varphi_i(\ub),
\qquad
g(\ub_{k+1}) \approx \sum_{i=1}^m \alpha_i \lambda_i \varphi_i(\ub_k).
\]
However, Koopman eigenfunctions are rarely available in closed form, which makes such a truncation impractical.

\paragraph{Finite-dimensional approximation via a dictionary of observables.}
Instead, one may build a finite-dimensional approximation using a user-chosen dictionary of observables.
Let $\{X_i\}_{i=1}^m \subset \Vc$ be a set of scalar observables and define the corresponding subspace
\[
\Vc_m := \mathrm{span}\{X_1,\dots,X_m\} \subset \Vc.
\]
This choice of $\Vc_m$ can be interpreted as selecting the class of observables that we will represent.
We collect the dictionary evaluations into the vector-valued observable map
\begin{equation}\label{eq:h_lift}
\mathbf{X}(\ub) :=
\begin{bmatrix}
X_1(\ub)\\ \vdots\\ X_m(\ub)
\end{bmatrix} \in \Rbb^m.
\end{equation}
Any observable $g \in \Vc_m$ can be expressed as a linear combination of the dictionary functions,
\begin{equation}\label{eq:g_in_span_h}
g(\ub) = \sum_{i=1}^m a_i X_i(\ub) = \mathbf a^\top \mathbf{X}(\ub),
\qquad \mathbf a \in \mathbb{C}^m.
\end{equation}

\begin{figure}
\centering
\includegraphics[width = 1.0\linewidth]{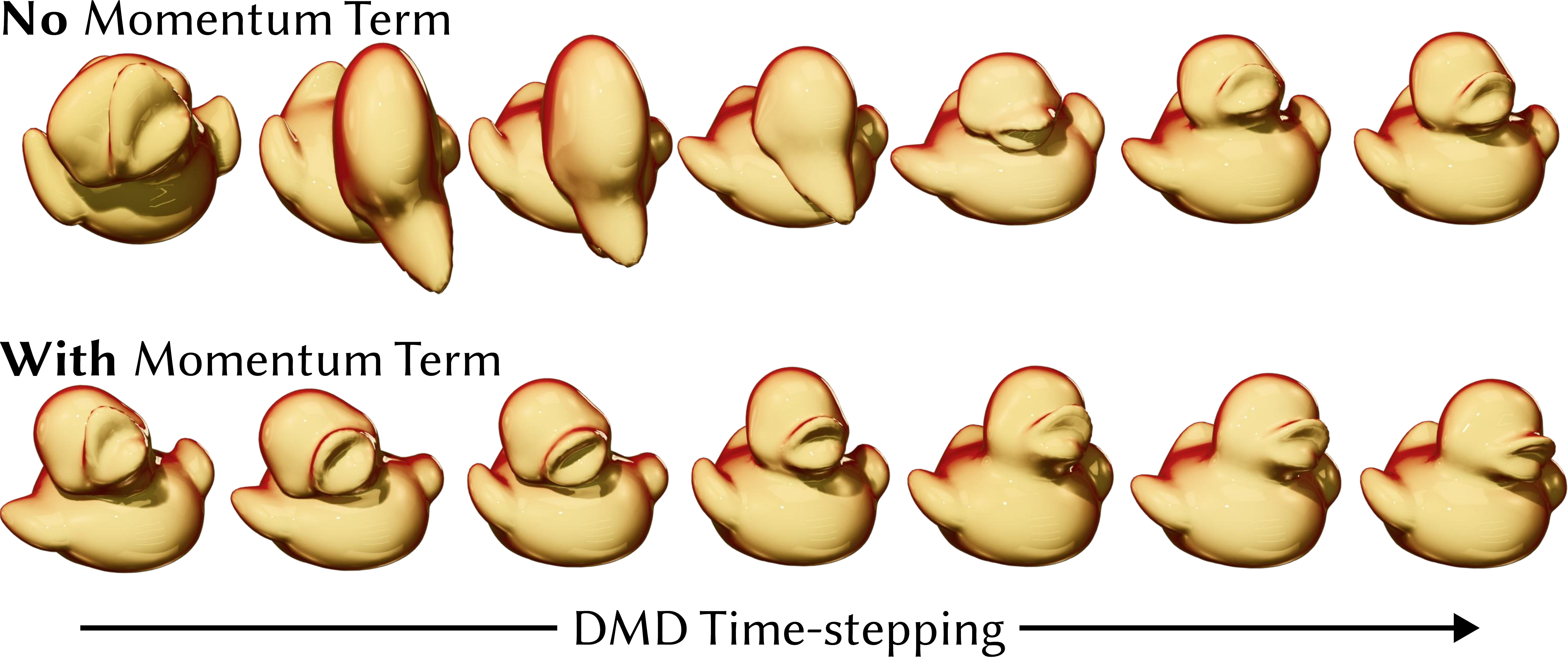}
\caption{\emph{Ablation on the momentum term.}
Unlike prior reduced-order models for deformable objects that represent the state using displacement alone, accurate operator approximation requires both displacement and momentum. 
We ablate the state representation accordingly: using displacement only leads to unstable and physically implausible behavior (top), while incorporating the momentum term yields realistic and stable dynamics (bottom).}
\label{fig:duck_momentum}
\end{figure}

The central modeling assumption behind Dynamic Mode Decomposition (DMD)~\cite{schmid2010dmd} is that the Koopman evolution closes (approximately) on this finite dictionary:
there exists a matrix $\mathbf{K} \in \Rbb^{m\times m}$ such that
\begin{equation}\label{eq:dmd_closure}
\mathbf{X}(\ub_{k+1})
= (\Kc \mathbf{X})(\ub_k)
= \mathbf{X}(\Fb(\ub_k))
\;\approx\;
\mathbf{K}\,\mathbf{X}(\ub_k),
\end{equation}
for all snapshot pairs in the dataset. In this way, the infinite-dimensional Koopman operator $\Kc$ is approximated by the
finite-dimensional linear map $\mathbf{K}$ acting on the coordinates induced by the chosen dictionary. Thus the dynamics can be approximated with 
\begin{equation}\label{eq:koopman}
    \mathbf{X}_{t+1} = \mathbf{K}\,\mathbf{X}_t
\end{equation}
with $\mathbf{X}_k : = \mathbf{X}( \ub_k )$. Note that, since $\mathbf{K}$ approximates Koopman only on $\Vc_m=\mathrm{span}\{X_i\}$, an inadequate choice of observables directly limits the fidelity of the recovered dynamics.

\paragraph{SVD-based Dynamic Mode Decomposition.}
Given snapshot data $\{\ub_k\}_{k=0}^{T}$, we form time-shifted data matrices in the lifted observable coordinates, namely
\begin{align*}
    \bm{\mathsf{X}}  &= [\,\mathbf{X}_{0},\, \mathbf{X}_{1},\, \ldots,\, \mathbf{X}_{T-1}\,], &
    \bm{\mathsf{X}}' &= [\,\mathbf{X}_{1},\, \mathbf{X}_{2},\, \ldots,\, \mathbf{X}_{T}\,],
\end{align*}
The closure assumption \eqref{eq:dmd_closure} over all snapshot pairs yields the matrix relation
\begin{equation}\label{eq:koopman_matrix}
\bm{\mathsf{X}}' \approx \mathbf{K}\,\bm{\mathsf{X}},
\end{equation}
and (E)DMD estimates $\mathbf{K}$ by solving the least-squares problem
\(
\min_{\mathbf{K}}\|\bm{\mathsf{X}}'-\mathbf{K}\bm{\mathsf{X}}\|_F.
\)
The minimum-norm solution is
\revision{\begin{equation}\label{eq:K_ls}
\mathbf{K} = \bm{\mathsf{X}}'\,\bm{\mathsf{X}}^{*}.
\end{equation}}

To compute this efficiently and robustly, we use the singular value decomposition (SVD)
\[
\bm{\mathsf{X}} = \mathbf{U}\boldsymbol{\Sigma}\mathbf{V}^\top,
\]
so that $\bm{\mathsf{X}}^* = \mathbf{V}\boldsymbol{\Sigma}^{-1}\mathbf{U}^\top$ and \eqref{eq:K_ls} becomes
\revision{
\begin{equation}\label{eq:K_svd}
    \mathbf{K} = \bm{\mathsf{X}}' \mathbf{V}\boldsymbol{\Sigma}^{-1}\mathbf{U}^\top.
\end{equation}
}
In practice, we use the truncated rank-$r$ SVD
$\bm{\mathsf{X}} \approx \mathbf{U}_r \boldsymbol{\Sigma}_r \mathbf{V}_r^\top$.
Since both $\bm{\mathsf{X}}$ and $\bm{\mathsf{X}}'$ lie approximately in the subspace spanned by~$\mathbf{U}_r$, 
we project $\mathbf{K}$ onto this subspace to obtain
\begin{equation}\label{eq:dmd_Kprime}
    \mathbf{K}' =  \mathbf{U}_r^\top \mathbf{K} \mathbf{U}_r 
    = \mathbf{U}_r^\top \bm{\mathsf{X}}' \mathbf{V}_r \boldsymbol{\Sigma}_r^{-1}
    \in \mathbb{R}^{r\times r},
\end{equation}
where $\mathbf{U}_r^\top \mathbf{U}_r = \mathbf{I}_r$.

We then perform eigenanalysis of the reduced operator,
\revision{
\begin{equation}
    \mathbf{K}' = \boldsymbol{\phi}\boldsymbol{\Lambda} \boldsymbol{\phi}^{-1} ,
\end{equation}
}
where $\boldsymbol{\phi}$ and $\boldsymbol{\Lambda}$ are the eigenvectors and eigenvalues in the reduced space.
The corresponding DMD modes in the full state space are obtained as
\begin{equation}
    \boldsymbol{\Phi} = \mathbf{U}_r \boldsymbol{\phi}.
\end{equation}
Finally, the full Koopman approximation can be expressed as
\begin{equation}
    \mathbf{K} \approx \boldsymbol{\Phi}\boldsymbol{\Lambda}\boldsymbol{\Phi}^{*}.
    \label{eq:low_rank_k}
\end{equation}

The eigenvalues and eigenvectors of DMD are generally complex-valued, 
since the reduced operator~$\mathbf{K}'$ is not necessarily symmetric. 
This complex spectrum allows DMD to capture oscillatory and nonlinear behavior in the trajectory, 
analogous to how exponentiating a complex scalar encodes rotation with a specific frequency.

\section{Adaptation to Deformable Simulation}

We adapt the DMD formulation for deformable simulations in graphics. 
Our objective is to construct a formulation that reproduces the qualitative behavior 
of traditional reduced-order models (ROMs) commonly used for real-time dynamics. 
Because DMD inherently builds a low-rank approximation of the system’s evolution, 
it serves as a natural counterpart to reduced-space simulation methods. 
We first develop this adaptation to achieve comparable interactive deformation performance, 
and later discuss the unique advantages introduced by this formulation.
\subsection{Time Integration}
As discussed in Equation~\ref{eq:koopman}, the system’s evolution can be represented as a linear propagation. 
Using the low-rank approximation of the Koopman operator in Equation~\ref{eq:low_rank_k}, 
the time stepping relation becomes
\begin{equation}
    \mathbf{X}_{t+1} = \boldsymbol{\Phi}\boldsymbol{\Lambda}\boldsymbol{\Phi}^{*}\mathbf{X}_t,
    \label{eq:time_stepping}
\end{equation}
where $\boldsymbol{\Phi}$ and $\boldsymbol{\Lambda}$ denote the DMD modes and eigenvalues, respectively as defined in Eq.~\eqref{eq:low_rank_k} .

\paragraph{Discussion: Connection to reduced-order time stepping.}
The update in Eq.~\eqref{eq:time_stepping} can be interpreted as a three-stage reduced-order time stepping procedure. 
Starting from the current observed state $\mathbf{X}_t$, multiplication by the pseudoinverse $\boldsymbol{\Phi}^*$ projects the state into the reduced Koopman subspace, analogous to computing reduced coordinates in classical model reduction. 
The diagonal operator $\boldsymbol{\Lambda}$ then advances these reduced coordinates forward in time, replacing a traditional time integrator with a linear propagation in the reduced space. 
Finally, multiplication by $\boldsymbol{\Phi}$ lifts the updated reduced state back to the full observable space, producing the next observed state $\mathbf{X}_{t+1}$.

Viewed this way, our formulation closely parallels reduced-order time stepping while replacing the usual numerical integrator with a learned Koopman operator that enables efficient temporal evolution.

\subsection{Modeling Momentum Term}

\begin{figure}
\centering
\includegraphics[width = 1.0\linewidth]{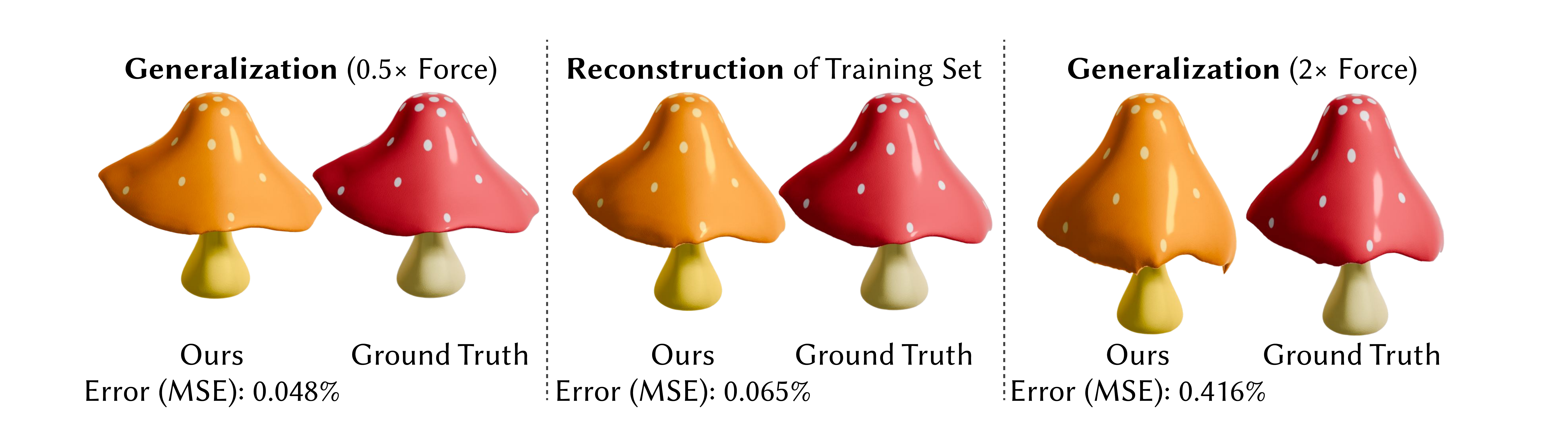}
\caption{\emph{Generalization to force magnitude.}
Our model is trained on a single force magnitude and evaluated on unseen forces scaled by $0.5\times$ and $2\times$. 
Despite these changes, the predicted deformations remain visually consistent with the ground truth and incur low reconstruction error, demonstrating that the learned Koopman dynamics generalize to variations in force magnitude beyond the training set.
}
\label{fig:mushroom_force_magnitude}
\end{figure}

\begin{figure*}
\centering
\includegraphics[width = 1.0\linewidth]{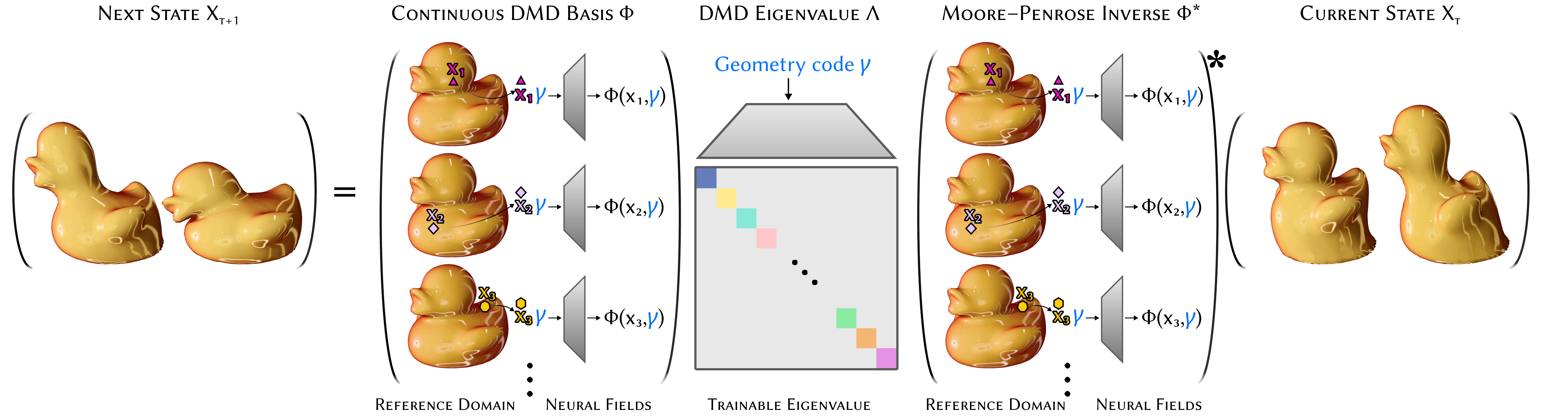}
\caption{\emph{Training pipeline.}
We represent the Koopman basis functions using neural fields, allowing them to be evaluated continuously over the reference domain and shared across different discretizations. 
Both the basis functions and the eigenvalues are further conditioned on a geometry code to support generalization across shapes. 
During training, the current state is projected into the Koopman subspace, advanced in time using the learned eigenvalues, and lifted back to predict the next state. 
The model is trained by minimizing the difference between this predicted next state and the ground-truth simulation.
}
\label{fig:training_pipeline}
\end{figure*}

As emphasized in previous sections, the choice of observables---equivalently, the finite-dimensional space $\Vc_m$ on which we approximate the Koopman operator---directly determines the fidelity and stability of the recovered dynamics.

In classical reduced-order simulation for elastodynamics, the reduced basis is typically constructed for the \emph{primary variable}, i.e., the displacement field $\mathbf{U}_t$~\cite{barbivc2005real, benchekroun2023FastComplemDynamics, Kim:Skipping:2009}. 
This is well-suited to traditional projection-based time integrators, which maintain an explicit notion of velocity and acceleration internally. 
However, DMD learns a \emph{first-order} evolution map $\mathbf{X}_{t+1} \approx \mathbf{K}\,\mathbf{X}_t$, and therefore requires that the observable $\mathbf{X}_t$ form a Markov state: it must contain all information needed to predict the next step.

For elastodynamics, displacement alone is not sufficient to satisfy this closure property.
Because the underlying equations are second-order in time, the evolution depends on both configuration and momentum (or velocity).
Consequently, choosing $\mathbf{X}_t=\mathbf{U}_t$ defines an observable space $\Vc_m$ that is too restrictive: there is no single linear operator that can reliably map displacement snapshots to the next displacement snapshot across the trajectory.
Empirically, as shown in Figure~\ref{fig:duck_momentum}, training DMD on $\mathbf{X}_t=\mathbf{U}_t$ leads to unstable or physically implausible rollouts.

To incorporate both deformation and momentum, we augment the state vector as
\begin{equation}
\mathbf{X}_t =
\begin{bmatrix}
\mathbf{U}_t \\
\mathbf{U}_t - \mathbf{U}_{t-1}
\end{bmatrix}
\in \mathbb{R}^{6n},
\label{eq:momentum_state}
\end{equation}
where $\mathbf{U}_t \in \mathbb{R}^{3n}$ stores vertex displacements, and 
$(\mathbf{U}_t - \mathbf{U}_{t-1})$ encodes their velocity components.
As illustrated in Figure~\ref{fig:duck_discretization_agnostic}, this formulation captures the system’s dynamics more accurately.

\paragraph{External forcing as an impulse in the lifted state.}
The Koopman/DMD operator constructed above models the autonomous evolution
$\mathbf{X}_{t+1}\approx \mathbf{K}\mathbf{X}_t$.
To incorporate time-varying external forces $\boldsymbol{f}_t := \boldsymbol{f}(t)$ during interactive simulation, where the dynamical systems takes the form $\ub_{t+1} = \Fb( \ub_t, \boldsymbol{f}_t)$, we treat forcing as an instantaneous update to the lifted state prior to applying the Koopman propagation. Here, $\boldsymbol{f}_t$ may include user-applied forces as well as other external forces. Concretely, we form the perturbed state
$\mathbf{X}_t^{+}=\mathbf{X}_t+\mathbf{F}_t$, where 
\begin{equation}
\mathbf{F} =
\begin{bmatrix}
\mathbf{0} \\
\mathbf{f}\,h^2
\end{bmatrix}
\in \mathbb{R}^{6n},
\label{eq:force}
\end{equation}
modifies only the velocity component, and then advance it using
the same operator:
\[
\mathbf{X}_{t+1}\approx \mathbf{K}\mathbf{X}_t^{+}= \mathbf{K}(\mathbf{X}_t+\mathbf{F}_t).
\]
By linearity this is equivalent to $\mathbf{X}_{t+1}\approx \mathbf{K}\mathbf{X}_t + \mathbf{K}\mathbf{F}_t$, i.e., a special case of a controlled
update (or DMDc ~\cite{proctor2016dmdc}) in which the input enters through the Koopman propagation itself. Moving forward, with perhaps some abuse in notation, we will not distinguish between $\mathbf{X}$ and $\mathbf{X}^+$. As shown in Figure~\ref{fig:mushroom_force_magnitude}, this force handling allows us to generalize to unseen forces.


\subsection{Lightning Fast Time Stepping}
\label{sec:property_temporal}
The linear approximation of the Koopman operator enables accelerated reduced-space time stepping that scales sub-linearly with respect to the number of time steps. 
Suppose we wish to advance the system by $N$ steps. 
The resulting state is
\begin{equation}
    \mathbf{X}_{t+N}
    = (\boldsymbol{\Phi}\boldsymbol{\Lambda}\boldsymbol{\Phi}^{*})^{N}\mathbf{X}_t
    = \boldsymbol{\Phi}\,\boldsymbol{\Lambda}^{N}\boldsymbol{\Phi}^{*}\mathbf{X}_t,
    \label{eq:time_stepping_N}
\end{equation}
where $\boldsymbol{\Lambda}$ is a diagonal matrix in the $r$-dimensional reduced space. 
Computing $\boldsymbol{\Lambda}^{N}$ requires exponentiating only its diagonal entries, 
yielding an $O(\log N)$ complexity with respect to the number of steps, 
in contrast to the $O(N)$ cost of conventional numerical integration. Another benefit of this reduction is that the matrix multiplication is performed entirely in the $r$-dimensional reduced space, further lowering the computational scaling of the integration step from $O(n)$ to $O(r)$.
A similar acceleration was demonstrated  for fluid simulation by~\citet{chen2025dmd}, 
but has not yet been explored for reduced-order deformable simulation in computer graphics.

This formulation also enables arbitrary adjustment of the simulation time-step size while preserving consistent dynamics by construction. 
Because the evolution operator acts through eigenvalues, changing the time step from the training value~$h$ to a new value~$h'$ amounts to rescaling these eigenvalues. 
If $\boldsymbol{\Lambda}(h)$ denotes the diagonal matrix associated with step size~$h$, 
then the operator corresponding to a new time step~$h'$ is
\begin{equation}
    \boldsymbol{\Lambda}(h')
    = \exp\!\left( \frac{h'}{h}\,\log \boldsymbol{\Lambda}(h) \right),
    \label{eq:timestep_rescaling}
\end{equation}
which correctly adjusts the temporal frequency and decay rates of each mode.
\revision{This formulation follows from viewing $\Lambda(h)$ as a discrete-time propagator: advancing the system by a step $h'$ corresponds to composing the original propagator $h'/h$ times, which in the continuous limit yields the exponential map above.}
As shown in Figure~\ref{fig:gummybear_comparison}, this rescaling gives consistent deformation behavior under different time step sizes, 
whereas conventional reduced-space integrators exhibit noticeably different dynamics when the time step is modified.

\else
\section{Background: Low-Rank Approximation of Koopman Operator}

We consider a physical system described by a state vector~$\mathbf{X}$, representing quantities such as displacement or velocity. 

Under Koopman theory, the system’s evolution can be expressed as a linear propagation 

\begin{equation}\label{eq:koopman}
    \mathbf{X}_{t+1} = \mathbf{K}\,\mathbf{X}_t,
\end{equation}
where~$\mathbf{K}$ is the (possibly infinite-dimensional) Koopman operator and~$t$ denotes the discrete time step; see~\cite{chen2025dmd} for a detailed discussion of its existence and properties.

Following Dynamic Mode Decomposition (DMD)~\cite{schmid2010dmd}, we estimate~$\mathbf{K}$ directly from simulation data. 
Given a time-shifted sequence of system states, we define
\begin{align*}
    \bm{\mathsf{X}}  &= [\,\mathbf{X}_{0},\, \mathbf{X}_{1},\, \ldots,\, \mathbf{X}_{T-1}\,], &
    \bm{\mathsf{X}}' &= [\,\mathbf{X}_{1},\, \mathbf{X}_{2},\, \ldots,\, \mathbf{X}_{T}\,],
\end{align*}
where each column~$\mathbf{X}_t$ stores the system state at time~$t$.
We seek a linear operator~$\mathbf{K}$ satisfying 
$\bm{\mathsf{X}}' \approx \mathbf{K}\bm{\mathsf{X}}$.
Using the singular value decomposition (SVD)
$\bm{\mathsf{X}} = \mathbf{U}\boldsymbol{\Sigma}\mathbf{V}^\top$,
we approximate
\begin{equation}
    \mathbf{K} \approx \bm{\mathsf{X}}' \mathbf{V}\boldsymbol{\Sigma}^{-1}\mathbf{U}^\top,
\end{equation}
where $\mathbf{V}\boldsymbol{\Sigma}^{-1}\mathbf{U}^\top$ is the pseudoinverse of~$\bm{\mathsf{X}}$.
In practice, we use the truncated rank-$r$ SVD
$\bm{\mathsf{X}} \approx \mathbf{U}_r \boldsymbol{\Sigma}_r \mathbf{V}_r^\top$.
Since both $\bm{\mathsf{X}}$ and $\bm{\mathsf{X}}'$ lie approximately in the subspace spanned by~$\mathbf{U}_r$, 
we project $\mathbf{K}$ onto this subspace to obtain
\begin{equation}\label{eq:dmd_Kprime}
    \mathbf{K}' =  \mathbf{U}_r^\top \mathbf{K} \mathbf{U}_r 
    = \mathbf{U}_r^\top \bm{\mathsf{X}}' \mathbf{V}_r \boldsymbol{\Sigma}_r^{-1}
    \in \mathbb{R}^{r\times r},
\end{equation}
where $\mathbf{U}_r^\top \mathbf{U}_r = \mathbf{I}_r$.

We then perform eigenanalysis of the reduced operator,
\begin{equation}
    \mathbf{K}' = \boldsymbol{\phi}\boldsymbol{\Lambda} \boldsymbol{\phi}^* ,
\end{equation}
where $\boldsymbol{\phi}$ and $\boldsymbol{\Lambda}$ are the eigenvectors and eigenvalues in the reduced space.
The corresponding DMD modes in the full state space are obtained as
\begin{equation}
    \boldsymbol{\Phi} = \mathbf{U}_r \boldsymbol{\phi}.
\end{equation}
Finally, the full Koopman approximation can be expressed as
\begin{equation}
    \mathbf{K} \approx \boldsymbol{\Phi}\boldsymbol{\Lambda}\boldsymbol{\Phi}^{*}.
    \label{eq:low_rank_k}
\end{equation}

The eigenvalues and eigenvectors of DMD are generally complex-valued, 
since the reduced operator~$\mathbf{K}'$ is not necessarily symmetric or positive definite. 
This complex spectrum allows DMD to capture oscillatory and nonlinear behavior in the trajectory, 
analogous to how exponentiating a complex scalar encodes rotation with a specific frequency.

\section{Adaptation to Deformable Simulation}

We adapt the DMD formulation for deformable simulations in graphics. 
Our objective is to construct a formulation that reproduces the qualitative behavior 
of traditional reduced-order models (ROMs) commonly used for real-time dynamics. 
Because DMD inherently builds a low-rank approximation of the system’s evolution, 
it serves as a natural counterpart to reduced-space simulation methods. 
We first develop this adaptation to achieve comparable interactive deformation performance, 
and later discuss the unique advantages introduced by this formulation.
\subsection{Time Integration}
As discussed in Equation~\ref{eq:koopman}, the system’s evolution can be represented as a linear propagation. 
Using the low-rank approximation of the Koopman operator in Equation~\ref{eq:low_rank_k}, 
the time-stepping relation becomes
\begin{equation}
    \mathbf{X}_{t+1} = \boldsymbol{\Phi}\boldsymbol{\Lambda}\boldsymbol{\Phi}^{*}\mathbf{X}_t,
    \label{eq:time_stepping}
\end{equation}
where $\boldsymbol{\Phi}$ and $\boldsymbol{\Lambda}$ denote the DMD modes and eigenvalues, respectively as defined in Eq.~\eqref{eq:low_rank_k} .

\paragraph{Discussion: Connection to reduced-order time stepping.}
The update in Eq.~\eqref{eq:time_stepping} can be interpreted as a three-stage reduced-order time-stepping procedure. 
Starting from the current state $\mathbf{X}_t$, multiplication by the pseudoinverse $\boldsymbol{\Phi}^*$ projects the state into the reduced Koopman subspace, analogous to computing reduced coordinates in classical model reduction. 
The diagonal operator $\boldsymbol{\Lambda}$ then advances these reduced coordinates forward in time, replacing a traditional time integrator with a linear propagation in the reduced space. 
Finally, multiplication by $\boldsymbol{\Phi}$ lifts the updated reduced state back to the full space, producing the next state $\mathbf{X}_{t+1}$.

Viewed this way, our formulation closely parallels reduced-order time stepping while replacing the usual numerical integrator with a learned Koopman operator that enables efficient temporal evolution.

\subsection{Modeling Momentum Term}

\begin{figure}
\centering
\includegraphics[width = 1.0\linewidth]{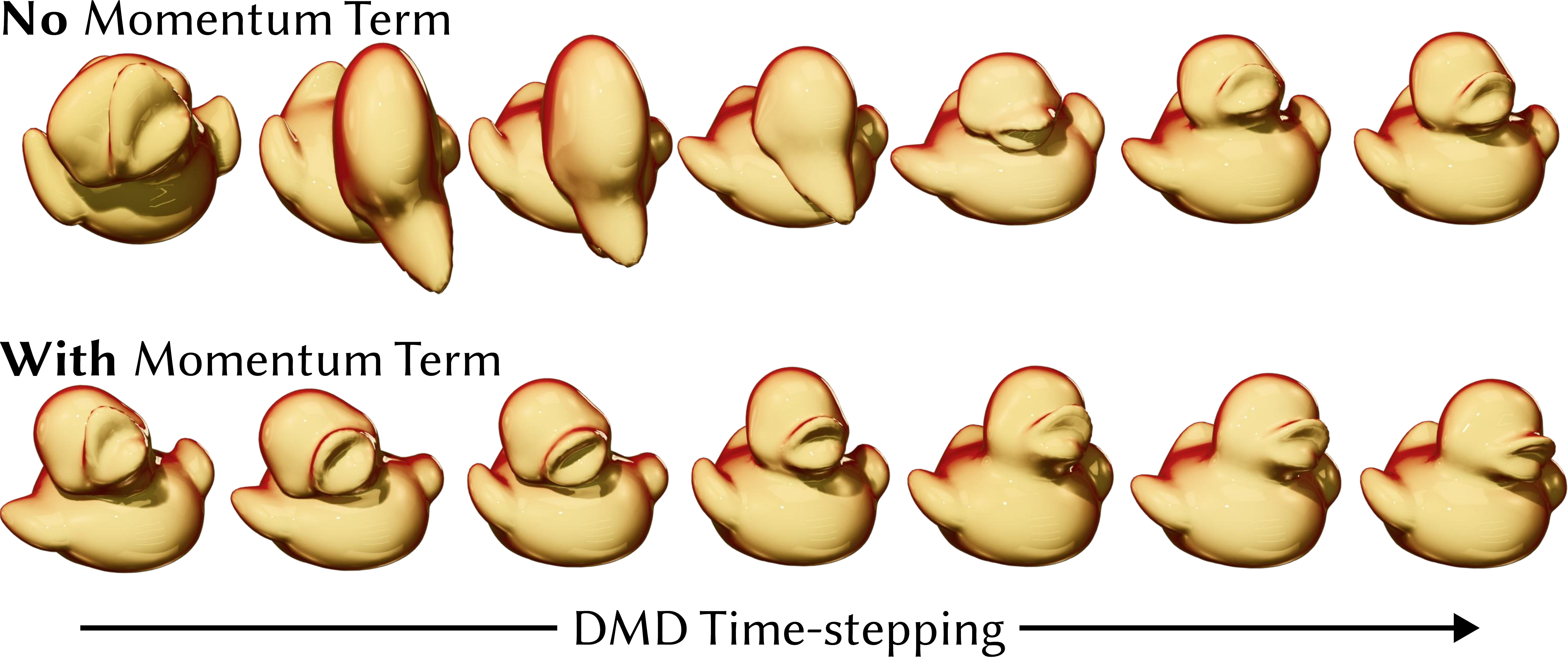}
\caption{\emph{Ablation on the momentum term.}
Unlike prior reduced-order models for deformable objects that represent the state using displacement alone, accurate operator approximation requires both displacement and momentum. 
We ablate the state representation accordingly: using displacement only leads to unstable and physically implausible behavior (top), while incorporating the momentum term yields realistic and stable dynamics (bottom).}
\label{fig:duck_momentum}
\end{figure}

An important consideration in constructing the system vector~$\mathbf{X}$ is how to represent dynamic information. 
Previous reduced-order deformable simulations typically define a reduced basis over the displacement field~\cite{barbivc2005real, benchekroun2023FastComplemDynamics, Kim:Skipping:2009}.

\todommc{ChangYue, did you try to use $(U_t, U_{t-1})$ or $(U_t, V_t \Delta t) $?}

This limitation arises because time stepping depends solely on the state variable: displacement alone does not encode the momentum necessary for dynamic evolution. 
To incorporate both deformation and momentum, we augment the state vector as
\begin{equation}
\mathbf{X}_t =
\begin{bmatrix}
\mathbf{U}_t \\
\mathbf{U}_t - \mathbf{U}_{t-1}
\end{bmatrix}
\in \mathbb{R}^{6n},
\label{eq:momentum_state}
\end{equation}
where $\mathbf{U}_t \in \mathbb{R}^{3n}$ stores vertex displacements, and 
$(\mathbf{U}_t - \mathbf{U}_{t-1})$ encodes their velocity components.
As illustrated in Figure~\ref{fig:duck_discretization_agnostic}, this formulation captures the system’s dynamics more accurately.

This construction resembles the second-order linear dynamical systems used by ~\citet{deAguiar2010Stable}, 
though their formulation is conditioned on input gestures, whereas our model addresses unconditioned deformable dynamics.

With this velocity-based formulation, external forces~$\mathbf{f}$ can be incorporated by adding their effect to the velocity component:
\begin{equation}
\mathbf{F} =
\begin{bmatrix}
\mathbf{0} \\
\mathbf{f}\,h^2
\end{bmatrix}
\in \mathbb{R}^{6n},
\label{eq:force}
\end{equation}
where $h$ is the time-step size. 
The system is then advanced in time under external forcing as
\begin{equation}
    \mathbf{X}_{t+1} = 
    \boldsymbol{\Phi}\boldsymbol{\Lambda}\boldsymbol{\Phi}^{*}
    (\mathbf{X}_t + \mathbf{F}_t),
    \label{eq:time_stepping_F}
\end{equation}
which integrates the applied force directly in the reduced Koopman update.


\subsection{Lightning Fast Time Stepping}
\label{sec:property_temporal}
The linear approximation of the Koopman operator enables accelerated reduced-space time stepping that scales sub-linearly with respect to the number of time steps. 
Suppose we wish to advance the system by $N$ steps. 
The resulting state is
\begin{equation}
    \mathbf{X}_{t+N}
    = (\boldsymbol{\Phi}\boldsymbol{\Lambda}\boldsymbol{\Phi}^{*})^{N}\mathbf{X}_t
    = \boldsymbol{\Phi}\,\boldsymbol{\Lambda}^{N}\boldsymbol{\Phi}^{*}\mathbf{X}_t,
    \label{eq:time_stepping_N}
\end{equation}
where $\boldsymbol{\Lambda}$ is a diagonal matrix in the $r$-dimensional reduced space. 
Computing $\boldsymbol{\Lambda}^{N}$ requires exponentiating only its diagonal entries, 
yielding an $O(\log N)$ complexity with respect to the number of steps, 
in contrast to the $O(N)$ cost of conventional numerical integration. 
A similar acceleration was demonstrated by~\citet{chen2025dmd} for fluid simulation, 
but has not yet been explored for reduced-order deformable simulation in computer graphics.

This formulation also enables arbitrary adjustment of the simulation time-step size while preserving consistent dynamics by construction. 
Because the evolution operator acts through eigenvalues, changing the time step from the training value~$h$ to a new value~$h'$ amounts to rescaling these eigenvalues. 
If $\boldsymbol{\Lambda}(h)$ denotes the diagonal matrix associated with step size~$h$, 
then the operator corresponding to a new time step~$h'$ is
\begin{equation}
    \boldsymbol{\Lambda}(h')
    = \exp\!\left( \frac{h'}{h}\,\log \boldsymbol{\Lambda}(h) \right),
    \label{eq:timestep_rescaling}
\end{equation}
which correctly adjusts the temporal frequency and decay rates of each mode. 
As shown in Figure~\ref{fig:gummybear_comparison}, this rescaling yields consistent deformation behavior under different time-step sizes, 
whereas conventional reduced-space integrators exhibit noticeably different dynamics when the time step is modified.

\fi

\section{Generalization Across Shapes}

The formulation of DMD introduced thus far is rooted in linear algebra, where both the system 
dimension and the basis functions are tied to a specific discretization. As in 
many reduced-order models, this dependence limits DMD to a single mesh or shape. 
Recent work in graphics has demonstrated that discretization-agnostic 
representations can lift this restriction, enabling reduced models to operate 
consistently across varying discretizations and geometries 
\cite{chang:2023:licrom, Modi:2024:Simplicits, chang2024neuralrepresentationshapedependentlaplacian}. 
Building on these ideas, we introduce a discretization-agnostic formulation of 
DMD, allowing a single temporally reduced model to generalize across multiple 
shapes and resolutions. Our training pipeline is shown in Figure~\ref{fig:training_pipeline}

\paragraph{Basis Representation}
Following this direction \revision{~\cite{chang2024neuralrepresentationshapedependentlaplacian}}, we represent each basis function 
$\bm{\phi}_i(\bm{x})$ in $\bm{\Phi} = [\bm{\phi}_1, \ldots, \bm{\phi}_r]$ as a 
continuous mapping over the rest configuration $\bm{x} \in \Omega$. Unlike 
previous work, which considers only real-valued vector fields, our formulation 
uses complex-valued basis functions 
$\bm{\phi}_i : \Omega \rightarrow \mathbb{C}^3$ in order to accommodate the 
complex modes that arise in DMD.

To support variation across geometries, we consider a family of domains 
$\{\Omega^{\gamma} \mid \gamma \in \mathcal{D}\}$ parameterized by a shape descriptor $\gamma$. 
Prior work has shown that basis functions can be conditioned on such a shape 
parameter, enabling $\bm{\phi}_i^{\,\gamma}(\bm{x}) \equiv \bm{\phi}_i(\gamma,\bm{x})$ to 
vary smoothly with both spatial position and underlying geometry 
\cite{chang2024neuralrepresentationshapedependentlaplacian}. We adopt this idea 
and extend it to the DMD setting by learning complex-valued, shape-conditioned 
modes $\bm{\phi}_i^{\gamma} : \Omega^{\gamma} \rightarrow \mathbb{C}^3$. This allows us 
to capture a smoothly varying family of complex spatiotemporal DMD modes across 
a continuous range of shapes.

\paragraph{Eigenvalue Representation}
Our formulation also requires a nonlinear mapping from shape code $\gamma$ to the eigenvalues $\boldsymbol{\Lambda}$, an $r$-dimensional 
complex-valued vector. Importantly, these eigenvalues depend only on the 
underlying geometry and not on the particular discretization. This is because 
the spectrum of the continuous operator—such as the deformation dynamics or its 
linearization—is an intrinsic property of the domain $\Omega^\gamma$. Different meshes 
merely approximate this operator at varying resolutions, but they do not change 
its eigenvalues in the continuous limit. Consequently, the eigenvalues vary only 
with the geometry descriptor $\gamma$, and we represent them as a geometry-conditioned 
function $\boldsymbol{\Lambda}^\gamma : \mathcal{D} \rightarrow \mathbb{C}^r$.

We parametrize both the basis functions $\bm{\phi}^{\gamma}$ and the eigenvalues $\boldsymbol{\Lambda}^\gamma$ using multilayer perceptrons (MLPs). The basis network $\bm{\phi}:\mathcal{D} \times \mathbb{R}^d \to \mathbb{C}^{3 \times r }$ takes as input a rest configuration point $\mathbf{x} \in \mathbb{R}^3$ from the reference domain $\Omega^\gamma$ along with the shape descriptor $\gamma$, and outputs a complex-valued matrix representing the local basis vectors. The eigenvalue network $\boldsymbol{\Lambda}: \mathcal{D} \to \mathbb{C}^r$ maps a shape descriptor $\gamma \in \mathcal{D}$ to a complex-valued vector of $r$ eigenvalues that parameterize the temporal evolution for the given geometry. \revision{Theoretically, $\bm{\phi}$ could be restricted to the reference domain $\Omega^\gamma$, but in practice, with the neural implementation, $\bm{\phi}$ is defined on all of $\mathbb{R}^3$. Only its values on the reference domain are physically meaningful, as the basis functions are only required to represent deformations originating from rest configurations in $\Omega^\gamma$.}


\subsection{Training}

We train $\bm{\phi}^{\gamma}$ and $\boldsymbol{\Lambda}^\gamma$ on simulation snapshots generated from full-space dynamics.

After collecting the time-shifted snapshots 
$\bm{\mathsf{X}} = [\mathbf{X}_{0}, \mathbf{X}_{1}, \ldots, \mathbf{X}_{T-1}]$ 
and 
$\bm{\mathsf{X}}' = [\mathbf{X}_{1}, \mathbf{X}_{2}, \ldots, \mathbf{X}_{T}]$, 
we optimize the network parameters by minimizing two losses. 
First, any deformation component that lies outside the span of the basis 
cannot be represented by the reduced model. To encourage the learned basis to 
capture the subspace underlying the data, we introduce a reconstruction loss 
that projects each deformation back onto the basis:
\begin{align}
\mathcal{L}_{\mathrm{recon}}
~=~
\sum_{t=0}^{T-1}
\bigl\|
\mathbf{X}_{t}
-
\boldsymbol{\Phi}^\gamma(\boldsymbol{\Phi}^\gamma)^{\!*}\mathbf{X}_{t}
\bigr\|_2^2 .
\label{loss:recon}
\end{align}
Here, $\boldsymbol{\Phi}^\gamma$ denotes the shape-conditioned basis for geometry 
$\gamma$, and $(\boldsymbol{\Phi}^\gamma)^{\!*}$ is implemented through a least-squares 
solve rather than an explicit pseudoinverse to avoid forming large dense 
matrices. This term enforces that each deformation $\mathbf{X}_t$ remains close 
to the subspace spanned by the learned basis; note that it does not encode any 
time-dependent information.

Second, to learn geometry-dependent eigenvalues that govern the dynamics in the 
reduced space, we introduce a time stepping loss that compares the predicted 
one-step evolution with the corresponding snapshot:
\begin{align}
\mathcal{L}_{\mathrm{step}}
~=~
\sum_{t=0}^{T-1}
\bigl\|
(\boldsymbol{\Phi}^\gamma)^{\!*}\mathbf{X}_{t+1}
-
\boldsymbol{\Lambda}^{\gamma}(\boldsymbol{\Phi}^\gamma)^{\!*}\mathbf{X}_{t}
\bigr\|_2^2 .
\label{loss:step}
\end{align}
As above, the operator $(\boldsymbol{\Phi}^\gamma)^{\!*}$ is computed via a 
least-squares projection. This term introduces the temporal structure of the 
data, training the eigenvalues $\boldsymbol{\Lambda}^\gamma$ to reproduce the 
correct one-step dynamics in the reduced coordinate space.

Finally, to ensure stable long-term behavior of the learned dynamics, we impose 
a constraint on the magnitude of each complex eigenvalue. Specifically, we 
regularize $\boldsymbol{\Lambda}^\gamma$ such that the combined norm of its real and 
imaginary parts satisfies
\begin{align}
|\lambda_i^\gamma|
~=~
\sqrt{(\operatorname{Re}\lambda_i^\gamma)^2 
      + (\operatorname{Im}\lambda_i^\gamma)^2}
~\le~ 1 ,
\qquad \forall i .
\label{loss:eig_constraint}
\end{align}
This mitigates the risk of unphysical exponential growth and promotes temporally bounded behavior aligned with conservative elastic dynamics. 


Specifically, after applying one time stepping operation using Eq.~\ref{eq:time_stepping}, the predicted deformation is required to match the corresponding frame in the training sequence. 
For our purposes, faithfully reproducing the induced dynamics is sufficient, and this behavior-oriented formulation proves effective in practice. 
As illustrated in Fig.~\ref{fig:duck_discretization_agnostic}, this enables the learned model to generalize across different mesh discretizations.

\revision{It is important to note that our training objective encourages the neural DMD bases and eigenvalues to reproduce the behavior of their classical linear-algebraic counterparts. Specifically, after applying one time stepping operation using Eq.~\ref{eq:time_stepping}, the predicted deformation is required to match the corresponding frame in the training sequence. However, the learned bases and eigenvalues are not constrained to match the exact eigenvectors or eigenvalues produced by standard linear algebra. While this differs from the standard DMD formulation, it remains consistent with the broader DMD family, where variants such as Extended DMD (EDMD) \cite{williams2015edmd} and Optimized DMD (OptDMD) \cite{askham2018variable} also produce distinct eigenfunction approximations. Our method similarly learns a low-rank linear approximation of the dynamics via supervised training Eq.~\ref{loss:step}, which is widely understood as a numerical approximation to the Koopman operator. For our purposes, faithfully reproducing the induced dynamics is sufficient, and this behavior-oriented formulation proves effective in practice. As illustrated in Fig.~\ref{fig:duck_discretization_agnostic}, this enables the learned model to generalize across different mesh discretizations.}

\begin{figure}
\centering
\includegraphics[width = 1.0\linewidth]{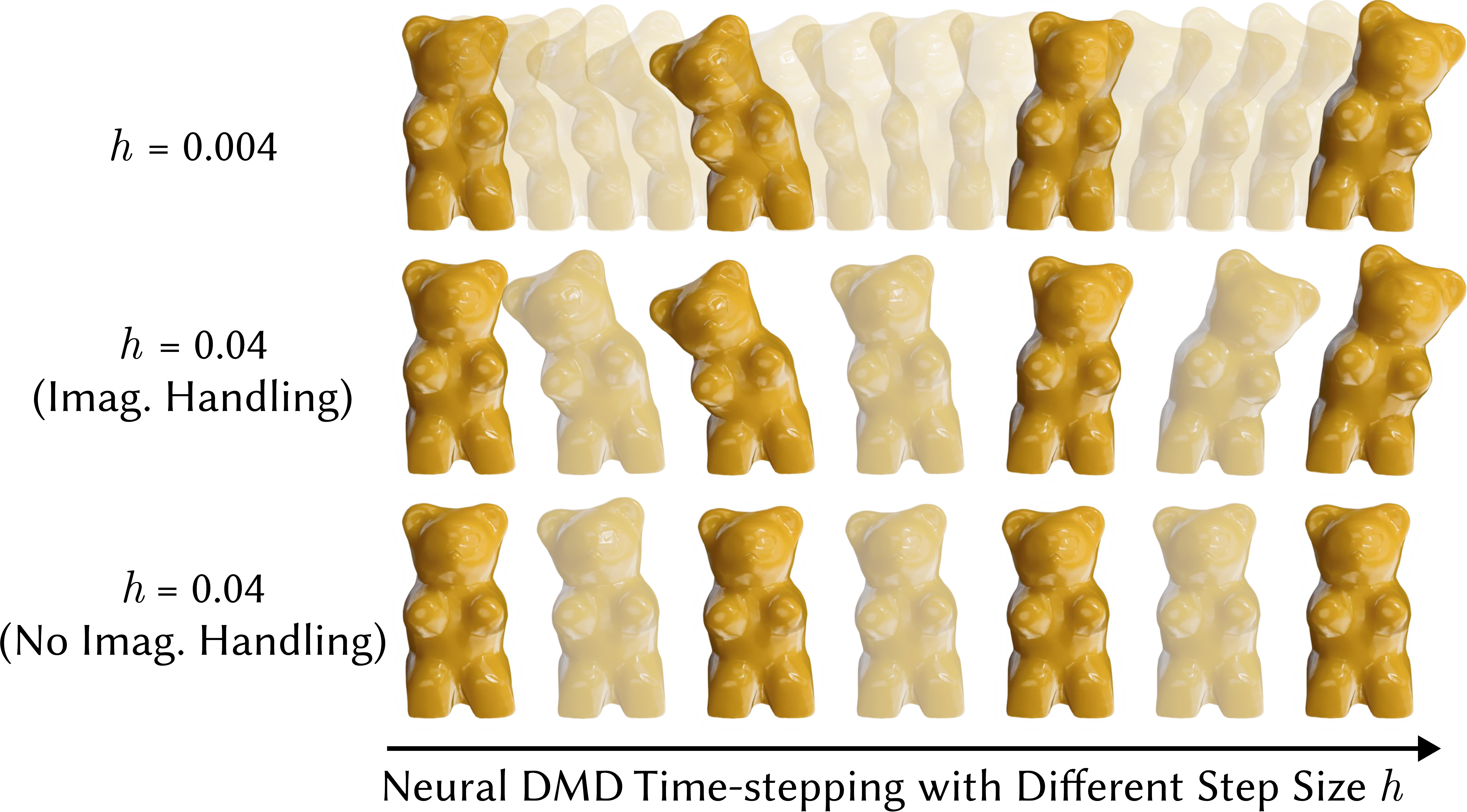}
\caption{\emph{Ablation on imaginary-part handling.}
Neural DMD time stepping is shown at two step sizes, $h=0.004$ and $h=0.04$. 
With imaginary-part projection enabled (middle), increasing the step size preserves the qualitative dynamics observed at smaller steps (top). 
Without imaginary-part handling (bottom), directly exponentiating the complex Koopman operator introduces spurious drift and altered behavior at larger time steps.
}
\label{fig:gummybear_ablation}
\end{figure}
\begin{figure*}
\centering
\includegraphics[width = 1.0\linewidth]{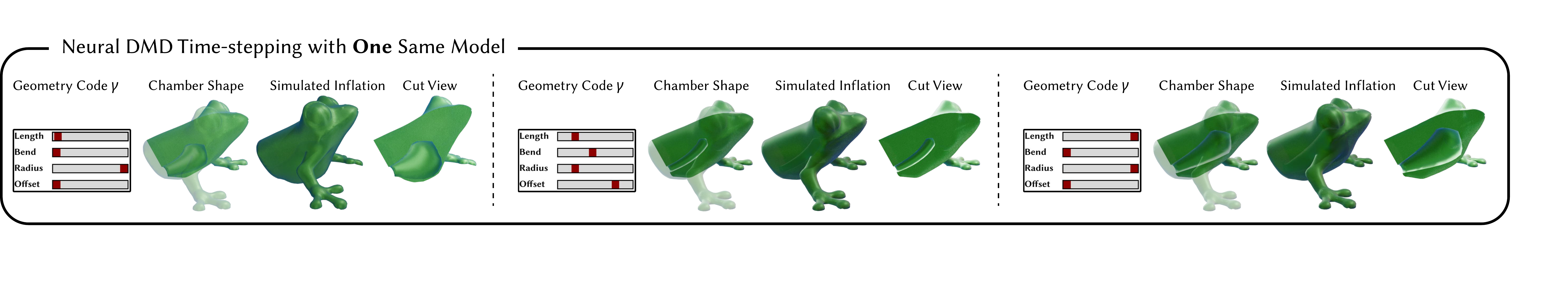}
\caption{\emph{Parameter-controlled simulation of inflatable frogs.} A single neural DMD model is trained over a family of frog-shaped chambers, where the geometry code $\gamma$ controls attributes such as chamber length, bend, radius, and offset. For each code $\gamma$, we visualize the resulting shape, simulated inflation, and internal chamber cutaway, all predicted using the same trained model.}
\label{fig:frog_shape_gen}
\end{figure*}

\subsection{Adaptation During Online Inference}

Although the observed state $\mathbf{X}_t$ is real-valued in our formulation, the Koopman factorization in Eq.~\ref{eq:time_stepping} involves complex-valued bases $\boldsymbol{\Phi}$ and eigenvalues $\boldsymbol{\Lambda}$. 
This implicitly embeds $\mathbf{X}_t$ into a complex latent space via $\boldsymbol{\Phi}^*\mathbf{X}_t$, which is benign when applying the operator one step at a time but can lead to bigger error when exponentiating the operator for long-horizon or large-step integration, as small imaginary components accumulate over time.

To ensure stable time stepping, we rewrite the Koopman operator in real form and explicitly project out the imaginary part for the physical state. 
Let $\boldsymbol{\Phi}=\mathbf{A}+i\mathbf{B}$ and $\boldsymbol{\Lambda}=\mathbf{C}+i\mathbf{D}$. 
We define their realified representations as
\begin{align*}
\boldsymbol{\Phi}_{\mathbb{R}}=
\begin{bmatrix}
\mathbf{A} & -\mathbf{B} \\
\mathbf{B} & \mathbf{A}
\end{bmatrix},
\qquad
\boldsymbol{\Lambda}_{\mathbb{R}}=
\begin{bmatrix}
\mathbf{C} & -\mathbf{D} \\
\mathbf{D} & \mathbf{C}
\end{bmatrix}.
\end{align*}
We then apply the projection
\begin{align*}
\mathbf{P}=
\begin{bmatrix}
\mathbf{I} & \mathbf{0} \\
\mathbf{0} & \mathbf{0}
\end{bmatrix},
\end{align*}
which removes all imaginary components in the physical space. 
Using this projection, we construct a low-dimensional real operator
\begin{align*}
\boldsymbol{K}_{\mathrm{real}}=\boldsymbol{\Phi}_{\mathbb{R}}^{*}\,\mathbf{P}\,\boldsymbol{\Phi}_{\mathbb{R}}\,\boldsymbol{\Lambda}_{\mathbb{R}} \in \mathbb{R}^{2r\times2r},
\end{align*}
where $r$ is the dimension of the reduced space.
Exponentiation of this operator produces stable real-valued time stepping without accumulation of imaginary drift. 
This post-processing step is particularly important in the neural setting, where learned eigenvalues and modes may contain small but non-negligible imaginary components, and enables reliable long-horizon and large-step integration.

In practice, this realified formulation is used only during time stepping. 
Given a real state $\mathbf{X}_t$, we first compute its complex reduced coordinates $\mathbf{z}_t=\boldsymbol{\Phi}^*\mathbf{X}_t$ and then convert them into their realified form $\mathbf{z}_{t,\mathbb{R}}=[\operatorname{Re}(\mathbf{z}_t)^\top,\ \operatorname{Im}(\mathbf{z}_t)^\top]^\top$. 
Time integration is then performed by applying the real Koopman operator $\boldsymbol{K}_{\mathrm{real}}$ (or its exponentiation) to $\mathbf{z}_{t,\mathbb{R}}$. 
The updated coordinates are finally converted back to complex form and lifted to the physical space via $\boldsymbol{\Phi}$ to produce the next real-valued state $\mathbf{X}_{t+1}$. 
This procedure ensures that imaginary components do not accumulate during long-horizon or large-step integration while preserving the expressiveness of the complex Koopman representation.

As shown in Fig.~\ref{fig:gummybear_ablation}, our imaginary-part handling yields consistent behavior even at larger time step sizes, whereas directly exponentiating the complex eigenvalues leads to noticeably different dynamics.




\section{Evaluations}

\revision{
We evaluate two implementations of the Koopman Deformables framework: a numerical variant (Figs.~
\ref{fig:teaser},
\ref{fig:mushroom_force_magnitude}, 
\ref{fig:armadillo_force_magnitude}, 
\ref{fig:flamingo_force_direction}, 
\ref{fig:duck_interactive}, 
\ref{fig:mouse}, and 
\ref{fig:finger_interactive}), which operates without shape-space modeling, and a neural variant, which enables it. 
For clarity, the neural results in Figs.  ~\ref{fig:duck_discretization_agnostic}, \ref{fig:gummybear_ablation}, and \ref{fig:real_world} are trained on a single shape with a fixed (zero) shape code, while shape-space generalization is demonstrated separately in Figs.~\ref{fig:frog_shape_gen} and \ref{fig:fringer_shape_gen}. 
For fixed boundary conditions (e.g., Figs.
\ref{fig:duck_discretization_agnostic},
\ref{fig:mushroom_force_magnitude}, 
\ref{fig:armadillo_force_magnitude}, 
\ref{fig:flamingo_force_direction}), constraints are enforced implicitly through the training data. By learning from trajectories that already satisfy these conditions, the basis encode the boundary behavior directly. As a result, no additional constraint handling is required at inference time, even when changing discretization or time step sizes.}

\revision{
Unless otherwise noted, all training sequences are generated using the open-source simulator of \cite{chang:2023:licrom}, which implements a stable Neo-Hookean elastic model~\cite{Smith:2018:stablenh} under fixed boundary conditions. These constraints are enforced by rewriting the displacements of pinned vertices to their target positions.
All results were obtained on an Intel i7-11800H CPU and an NVIDIA RTX 3070 GPU with 8~GB of memory.
}

\begin{table}[ht]
\centering
\rowcolors{2}{white}{gray!20} 
\begin{tabularx}{\linewidth}{l |>{\centering\arraybackslash}p{1.0cm} |>{\centering\arraybackslash}X |>{\centering\arraybackslash}X |>{\centering\arraybackslash}X}
 & \textbf{\# basis} & \textbf{\# vertices} & \textbf{\# training snapshots} & \textbf{Precompute time}  \\
\textbf{Figure~\ref{fig:teaser}} & 50 & 15k & 594 & 10.74 s  \\
\textbf{Figure~\ref{fig:mushroom_force_magnitude}} & 50 & 10k & 73 & 0.63 s  \\
\textbf{Figure~\ref{fig:armadillo_force_magnitude}} & 50 & 11k & 119 & 1.77 s  \\
\textbf{Figure~\ref{fig:flamingo_force_direction}} & 25 & 10k & 720 & 9.55 s  \\
\textbf{Figure~\ref{fig:duck_interactive}} & 50 & 10k & 392 & 3.78 s  \\
\textbf{Figure~\ref{fig:mouse}} & 50 & 10k & 544 & 6.81 s  \\
\textbf{Figure~\ref{fig:finger_error}} & 50 & 10k & 585 & 6.44 s  \\
\textbf{Figure~\ref{fig:duck_discretization_agnostic}} & 20 & 2.5k-500k & 392 &  7.33 min \\
\textbf{Figure~\ref{fig:gummybear_ablation}} & 20 & 10k & 248 &  1.33 h \\
\textbf{Figure~\ref{fig:frog_shape_gen}} & 20 & 10k & 3552 &  1.10 h \\
\textbf{Figure~\ref{fig:fringer_shape_gen}} & 20 & 10k & 17632 &   2.93 h\\
\textbf{Figure~\ref{fig:real_world}} & 5 & 68-70 & 94-135 &  11.42 min \\
\end{tabularx}
\caption{\revision{Summary of training snapshots and precomputation time. We report the number of basis functions, vertices, and training snapshots for each example. Figures ~\ref{fig:duck_discretization_agnostic}, \ref{fig:gummybear_ablation}, \ref{fig:frog_shape_gen}, \ref{fig:fringer_shape_gen}, and \ref{fig:real_world} use the neural implementation, while the remaining examples use the numerical formulation.}}
\label{tab:sequence_table}
\end{table}

\begin{table}[ht]
\centering
\rowcolors{2}{white}{gray!20} 
\begin{tabularx}{\linewidth}{l |>{\centering\arraybackslash}X |>{\centering\arraybackslash}X |>{\centering\arraybackslash}X |>{\centering\arraybackslash}X}
 & \textbf{Time/step (ours)} & \textbf{Time/seq (ours)} & \textbf{Time/step (full)} & \textbf{Time/seq (full)} \\
\textbf{Figure~\ref{fig:teaser}} & 4 ms & 11 ms & 711 ms & 7.6 min  \\
\textbf{Figure~\ref{fig:mushroom_force_magnitude}} & 8 ms & 10 ms & 1493 ms & 5.5 min  \\
\textbf{Figure~\ref{fig:armadillo_force_magnitude}} & 6 ms & 18 ms & 1408 ms &  6.9 min \\
\textbf{Figure~\ref{fig:flamingo_force_direction}} & 3 ms & 13 ms & 1520 ms &  6.2 min \\
\textbf{Figure~\ref{fig:duck_interactive}} & 4 ms & 10 ms & 542 ms & 1.8 min  \\
\textbf{Figure~\ref{fig:mouse}} & 2 ms & 15 ms  & 524 ms &  13.1 min \\
\textbf{Figure~\ref{fig:finger_error}} & 49 ms & 55 ms & 3018ms &  17.9 min \\
\textbf{Figure~\ref{fig:duck_discretization_agnostic}} & 4 ms & 7 ms & 542 ms & 1.8 min  \\
\textbf{Figure~\ref{fig:gummybear_ablation}} & 2 ms & 11 ms & 334 ms & 2.8 min  \\
\textbf{Figure~\ref{fig:frog_shape_gen}} & 11 ms & 15 ms & 541 ms & 7.21 min  \\
\textbf{Figure~\ref{fig:fringer_shape_gen}} & 3 ms & 9 ms & 536 ms & 13.4 min  
\end{tabularx}
\caption{\revision{Runtime comparison. \textbf{Time/step} measures per-step cost, while \textbf{Time/seq} reports the total time for a full simulation sequence. Our method is faster per step and, through temporal reduction, can take large steps to reach the final frame, yielding orders-of-magnitude speedups over full-space simulation.}}
\label{tab:timing_table}
\end{table}

\subsection{Time Step Invariance Compared to Implicit Euler}

Figure~\ref{fig:gummybear_comparison} compares our DMD time stepping against implicit Euler at two step sizes, $h=0.004$ and $h=0.04$, using both numerical and neural implementations. 
For implicit Euler, increasing the step size introduces strong numerical damping: the deformation quickly loses oscillatory motion and the kinetic energy decays much more rapidly, producing qualitatively different dynamics. 
This behavior is a well-known property of implicit integrators applied to stiff deformable systems, and it becomes more pronounced as the time step grows.

In contrast, both our numerical and neural formulations exhibit near-invariant behavior across step sizes. 
As shown in the middle and right columns of Fig.~\ref{fig:gummybear_comparison}, increasing $h$ preserves both the qualitative deformation patterns and the temporal structure of the kinetic energy, even when the step size is increased by an order of magnitude. 
This is a direct consequence of representing the dynamics with a linear Koopman operator: larger time steps correspond to exponentiating this operator, rather than repeatedly applying a numerical integrator.

Crucially, this also leads to improved computational efficiency. 
Exponentiating a low-rank operator scales only logarithmically with the number of time steps, whereas implicit Euler requires a Newton solve at every step, leading to linear scaling in time. 
As a result, our formulation provides both improved temporal fidelity at large step sizes and substantially faster long-horizon simulation, which is particularly important for interactive and optimization-driven applications.


\begin{figure}
\centering
\includegraphics[width = 1.0\linewidth]{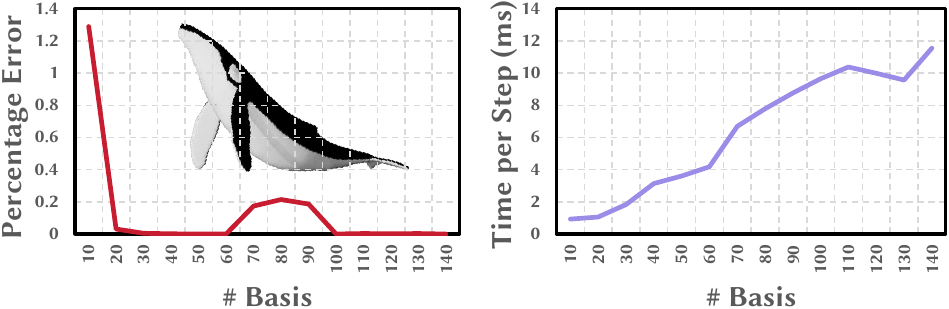}
\caption{\emph{Error and timing versus basis size $r$.} \revision{The prediction error decreases rapidly for small $r$, then becomes mildly non-monotonic at larger $r$ due to noise amplification and projection effects, while remaining low overall. Computation time increases with $r$ as the reduced operator grows, but stays within milliseconds.}}
\label{fig:whale_error}
\end{figure}

\subsection{Effect of Basis Size on Accuracy and Runtime}
\revision{
Fig.~\ref{fig:whale_error} evaluates the effect of truncation rank $r$ on prediction accuracy and runtime. Increasing $r$ initially reduces the one-step prediction error rapidly, confirming that the dominant dynamics are captured by a small number of modes. Beyond this regime, the error exhibits mild non-monotonic behavior while remaining low overall. This is expected: truncated SVD is optimal for reconstructing $\mathbf{X}$, but not for approximating its pseudoinverse in Eq.~\ref{eq:K_svd}. Since DMD estimates the operator via $\mathbf{X}^*$, inclusion of modes associated with small singular values introduces ill-conditioning and amplifies errors, leading to a bias--variance trade-off rather than monotonic improvement.
}

\revision{
Computation time increases with $r$ due to the larger reduced operator and eigenvalue system. However, all timings remain within milliseconds, indicating that even moderately large bases preserve real-time performance.
}

\subsection{Generalization}

\begin{figure}
\centering
\includegraphics[width = 1.0\linewidth]{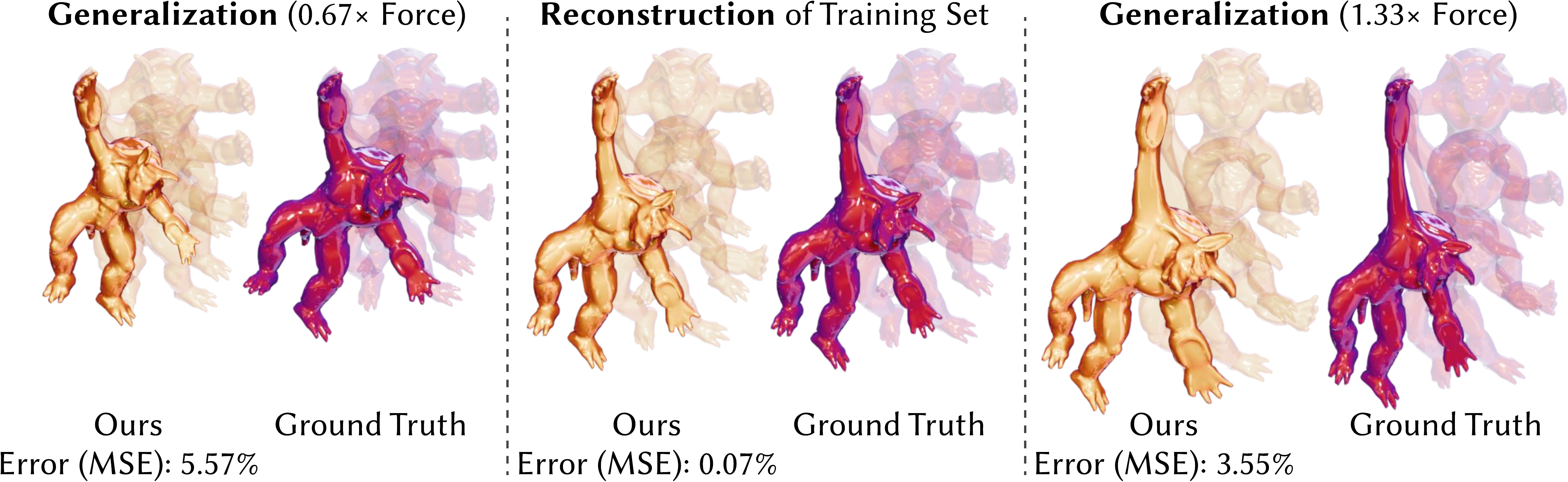}
\caption{\emph{Generalization to force magnitude on nonlinear trajectories.}
The model is trained on a single force magnitude and evaluated on unseen forces scaled by $0.67\times$ and $1.33\times$. 
Although the resulting trajectories are highly nonlinear, our method produces deformations that remain visually consistent with the ground truth and achieve low reconstruction error, demonstrating robust generalization beyond the training conditions.
}
\label{fig:armadillo_force_magnitude}
\end{figure}

\begin{figure}
\centering
\includegraphics[width = 1.0\linewidth]{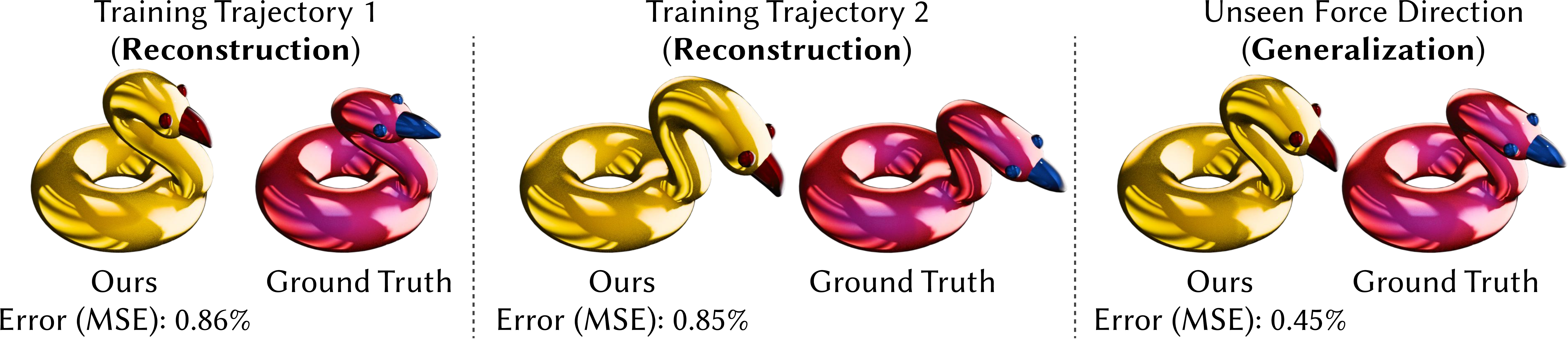}
\caption{\emph{Generalization to force direction.}
The model is trained on two force directions (left and center) and evaluated on an unseen direction (right). 
Despite this change, our method produces deformations that closely match the ground truth, demonstrating that the learned dynamics generalize to force directions not observed during training.
}
\label{fig:flamingo_force_direction}
\end{figure}

\begin{figure*}
\centering
\includegraphics[width = 1.0\linewidth]{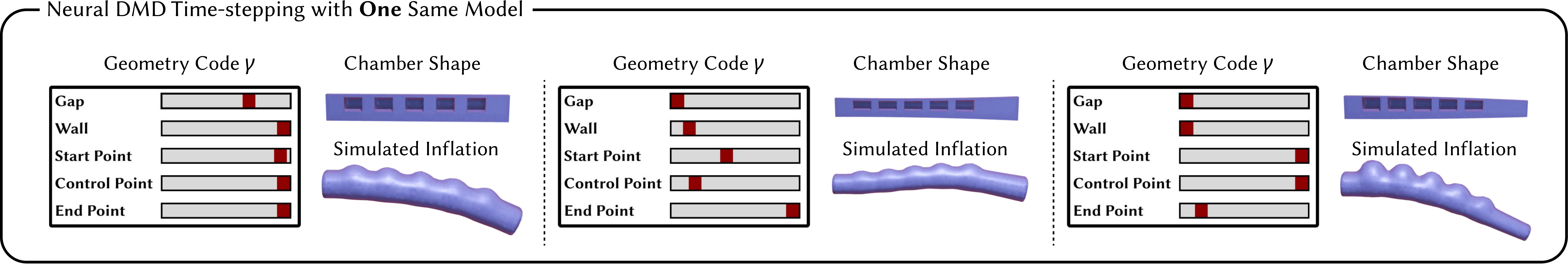}
\caption{\emph{Simulation of soft robotic fingers with parameterized chamber designs.} We use the same neural DMD model to simulate the inflation response of multiple finger designs, each defined by a different geometry code $\gamma$ (e.g., wall thickness, chamber gap, and Bézier control point). The resulting deformations reflect the effect of varying design parameters.}
\label{fig:fringer_shape_gen}
\end{figure*}

We evaluate generalization along two axes: \textbf{force}, and \textbf{shape}. 
In all experiments, we report the percentage mean squared error (MSE) between our results and full-space simulations, averaged over all frames of each trajectory.

\emph{Force generalization.}
We first evaluate robustness to force magnitudes not seen during training. 
Using our adapted numerical DMD formulation trained on a full-space simulation of a mushroom deforming under gravity (Figure~\ref{fig:mushroom_force_magnitude}), we test force magnitudes scaled to $0.5\times$ and $2\times$ the training value. 
Despite these unseen conditions, the resulting deformations remain visually consistent with the ground truth, with MSE below $0.5\%$. This behavior persists even for highly nonlinear deformations. 
For the armadillo example (Figure~\ref{fig:armadillo_force_magnitude}), our linearized dynamics accurately reconstruct and extrapolate to force magnitudes of $2/3\times$ and $4/3\times$ the training force, producing visually plausible motion. Finally, we evaluate generalization to unseen force directions. 
As shown in Figure~\ref{fig:flamingo_force_direction}, force directions outside the training set yield deformations that closely match the full-space simulation both qualitatively and quantitatively.

\emph{Discretization and shape generalization.}
Our discretization-agnostic extension of DMD introduces an additional axis of generalization: robustness across different mesh discretizations and geometric variations. As shown in Figure~\ref{fig:duck_discretization_agnostic}, a single trained neural model produces consistent DMD time stepping behavior across markedly different discretizations, ranging from $2.5$k to $10$k and up to $500$k vertices. This level of consistency across resolutions is difficult to achieve with traditional numerical methods.

Moreover, we are able to generalize across different shapes in addition to discretizations. Since our neural network is trained to represent the deformation behavior of parameterized domains $\Omega^{\gamma}$, where $\gamma$ encodes geometry, a single neural model can reproduce the deformation and physical response of an entire shape family. We show two examples of parameterized domains $\Omega^{\gamma}$ and their corresponding simulated deformations in Figure~\ref{fig:frog_shape_gen} and Figure~\ref{fig:fringer_shape_gen}. In both cases, our model qualitatively captures the geometry-dependent deformation induced by varying chamber shapes and placements.

\section{Applications}

\begin{figure}
\centering
\includegraphics[width = 1.0\linewidth]{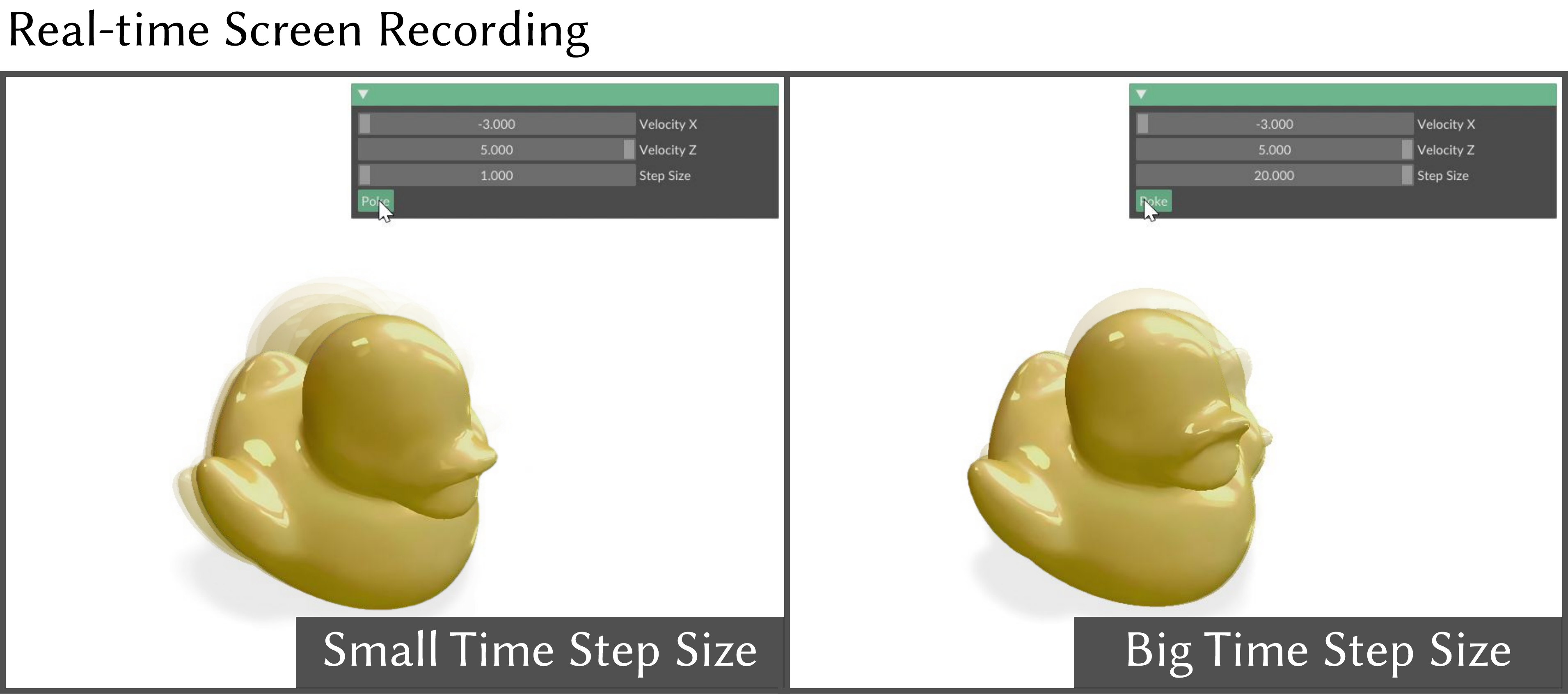}
\caption{\emph{Real-time interactive simulation.}
A user interactively deforms the duck model in real time by applying forces through a graphical interface. 
Our formulation allows the time step size to be adjusted during interaction, enabling flexible and responsive real-time manipulation.
}
\label{fig:duck_interactive}
\end{figure}

\begin{figure}
\centering
\includegraphics[width = 1.0\linewidth]{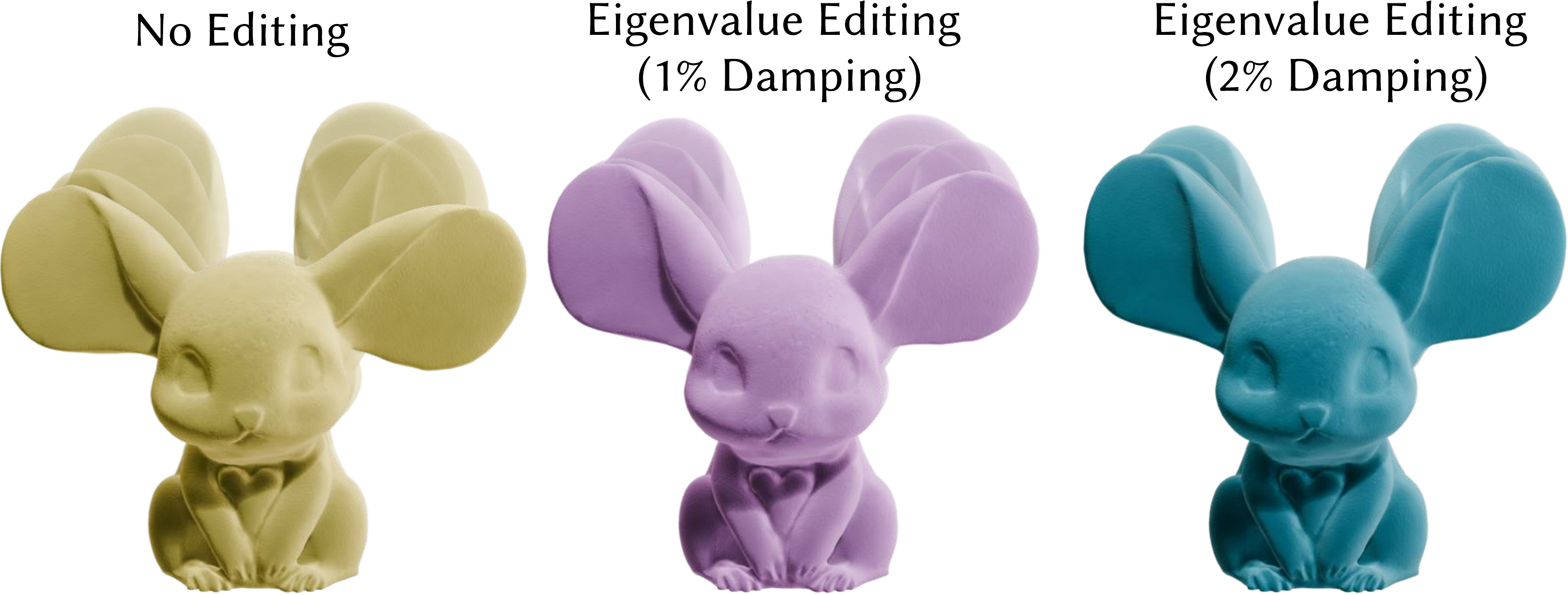}
\caption{\emph{Physical Behavior Editing via Eigenvalue Manipulation.} 
    By scaling the eigenvalues of the Koopman operator, we introduce global damping effects without retraining the model. 
    From left to right: original dynamics with no editing, 1\% damping, and 2\% damping. 
    All results are produced using the same learned model, with real-time performance.}
\label{fig:mouse}
\end{figure}

\subsection{Real-time Interactive Simulation}
Similar to prior reduced-order modeling approaches, our method also supports real-time interactive simulation. As shown in Figure ~\ref{fig:duck_interactive}, we demonstrate interactive manipulation of an elastic duck, where the user can freely drag, poke, and deform the model in real time. Thanks to the strong generalization of our learned reduced model, the system can robustly respond to forces and directions not present in the training set: the user can continuously vary both the force direction and magnitude, and the model adapts smoothly.

\subsection{Physical Behavior Editing via Eigenvalue Manipulation.}
Beyond real-time interaction, our formulation also supports direct editing of physical behavior through simple modifications to the Koopman operator. 
Since the eigenvalues of the reduced linear operator $\mathbf{K}$ characterize the temporal behavior of deformation modes, modifying their magnitude effectively changes the dynamic response of the system.

In particular, damping can be introduced by scaling each eigenvalue $\boldsymbol{\Lambda}$ by a factor $\mu \in (0,1)$:
\begin{align*}
\tilde{\boldsymbol{\Lambda}} = (1-\mu) \cdot \boldsymbol{\Lambda}.
\end{align*}
This directly alters the rate at which oscillations decay over time without retraining. 

As shown in Figure~\ref{fig:mouse}, we demonstrate this capability by applying increasing damping to a simulated elastic mouse. 
Without editing, the oscillations persist; with 1\% and 2\% damping applied via eigenvalue scaling, the motion attenuates visibly faster.

\subsection{Interactive Control via Fast Time Stepping}
\label{sec:iteractive_control}

\begin{figure}
\centering
\includegraphics[width = 1.0\linewidth]{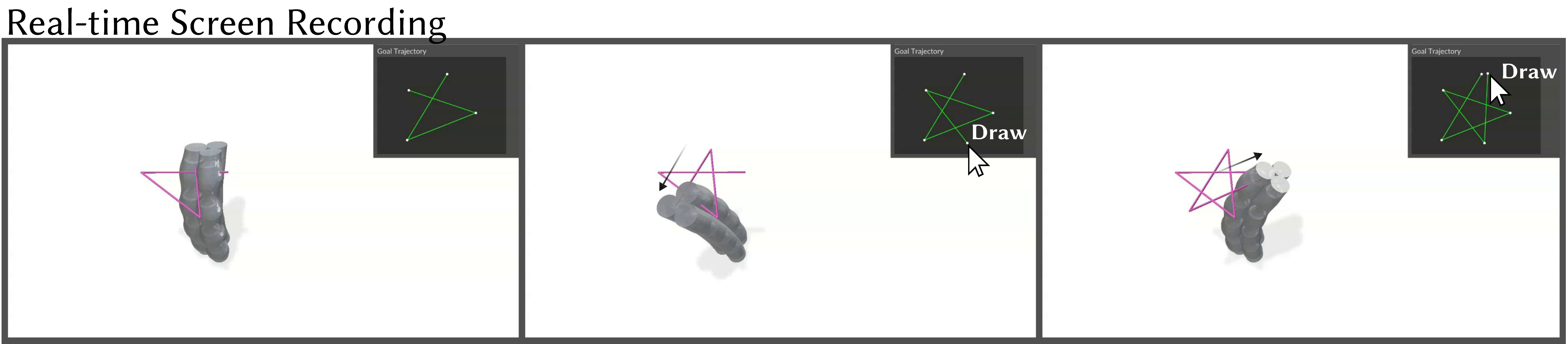}
\caption{\emph{Interactive control of a soft pneumatic finger.}
The user specifies a target trajectory by drawing in a 2D interface. 
Using fast Koopman time stepping, the system interactively solves for the internal air pressure that drives the finger to follow the drawn motion in real time.
}
\label{fig:finger_interactive}
\end{figure}

\begin{figure}
\centering
\includegraphics[width = 1.0\linewidth]{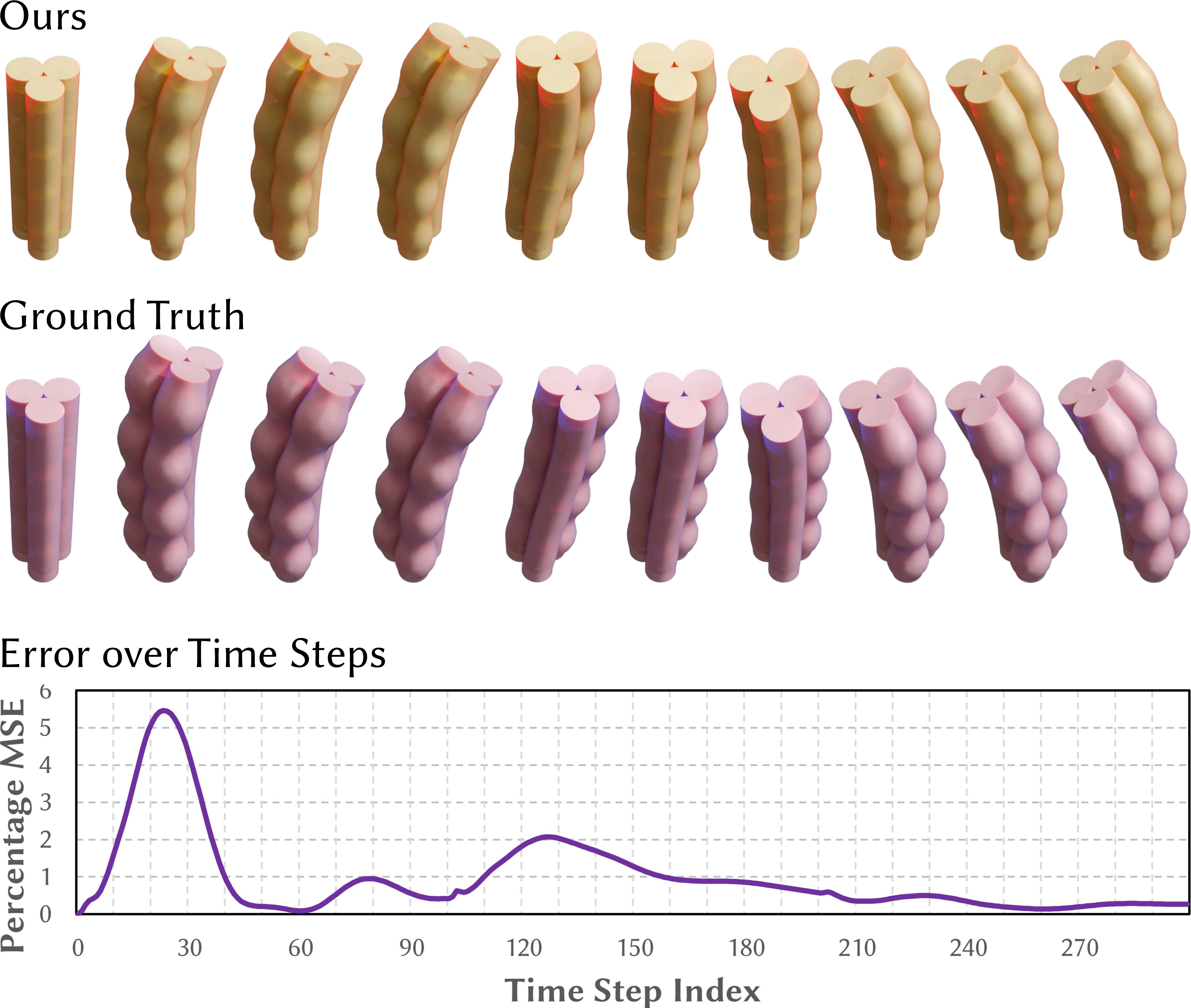}
\caption{\emph{Evaluation of control accuracy.}
The pressure signal computed by our method is applied to both the reduced model (top) and a full-space simulation (middle). 
The resulting deformations closely match over time, with the per-frame percentage MSE remaining below 2\% for most frames and below 6\% for all frames (bottom), demonstrating accurate and stable control under nonlinear pneumatic actuation.
}
\label{fig:finger_error}
\end{figure}

\begin{figure*}
\centering
\includegraphics[width = 1.0\linewidth]{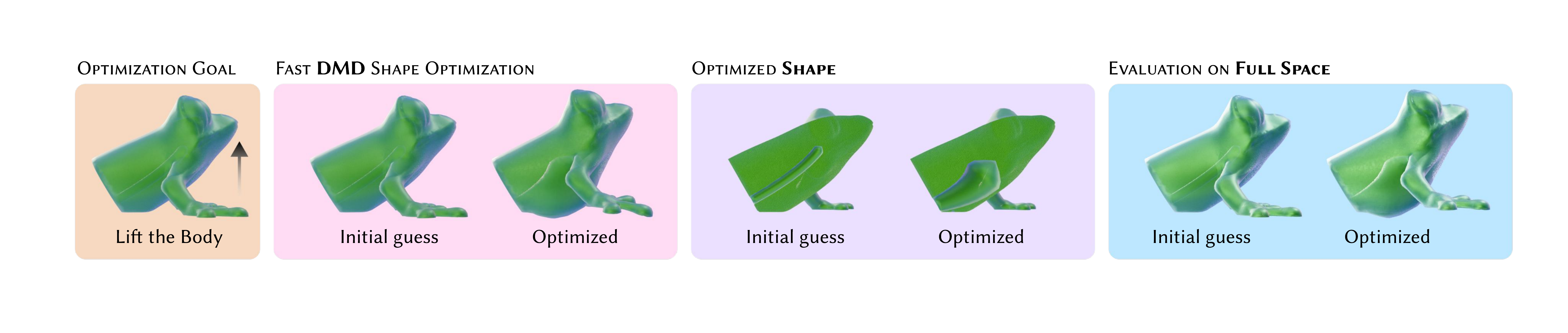}
\caption{\emph{Fast shape optimization with neural DMD.} Left: the optimization objective is to maximize vertical lift of the frog’s head region under pneumatic actuation. Middle: using fast Koopman time stepping, we optimize the internal chamber shape to achieve larger upward motion of the head. Right: full-space simulation of the optimized geometry confirms qualitatively similar behavior, validating the reduced-model optimization.}
\label{fig:frog_shape_opt}
\end{figure*}

The log-linear time stepping enabled by our formulation makes inverse problems such as control-force estimation tractable at interactive rates. 
We demonstrate this capability on a soft pneumatic robot finger, where the internal air pressure serves as the control input. 
Pressure is modeled as an explicit force applied to the faces of the interior chamber, and the resulting deformation is governed by the learned DMD dynamics.

The control variable $\mathbf{C} \in \mathbb{R}^k$ represents the pressure magnitudes applied to $k$ inner chambers. The corresponding force is modeled as
\[
\mathbf{F}_{\text{pressure}} = \mathbf{A}(\mathbf{X})\, \mathbf{C},
\]
where $\mathbf{A}(\mathbf{X})$ encodes how the current deformed state $\mathbf{X}$ translates pressure into force—e.g., based on deformed triangle areas on the internal surface. \revision{Following \cite{matyka2021pressure}, we distribute pressure forces to vertices proportionally to the deformed areas of adjacent faces, with directions given by the corresponding face normals.} Since $\mathbf{A}(\mathbf{X})$ depends on $\mathbf{X}$, computing the final state by traditional simulation would require evaluating the pressure-induced force at every time step, making the total cost linear in the number of steps $N$.

Instead, we assume the system reaches a quasi-static state with negligible velocity, allowing us to assume a constant force at final state. Under this assumption, and assuming a fixed linear Koopman operator during each iteration, the final deformation due to pressure can be approximated as:
\begin{equation}
\label{eq:air_final}
\mathbf{X}_{\text{final}} \approx \boldsymbol{\Phi} \left( \sum_{t=1}^N \boldsymbol{\Lambda}^t \right) \boldsymbol{\Phi}^* \mathbf{A}(\mathbf{X})\, \mathbf{C}.
\end{equation}
The term $\sum_{t=1}^N \boldsymbol{\Lambda}^t$ acts as a time-integrated propagator in the reduced space, and can be computed efficiently: in the diagonal case, it has a closed-form solution, and for general (non-diagonal) $\boldsymbol{\Lambda}$, we compute it via repeated squaring in $O(\log N)$ time.

Given this expression, we can solve for $\mathbf{C}$ using a least-squares formulation. Suppose we prescribe the target on a subset of degrees of freedom via a linear selection matrix $\mathbf{S}$, such that $\mathbf{S} \mathbf{X}_{\text{final}} = \mathbf{X}_{\text{goal}}$. Then we solve:
\begin{align*}
\mathbf{C}^* = \arg\min_{\mathbf{C}} \left\| \mathbf{S} \boldsymbol{\Phi} \left( \sum_{t=1}^N \boldsymbol{\Lambda}^t \right) \boldsymbol{\Phi}^* \mathbf{A}(\mathbf{X})\, \mathbf{C} - \mathbf{X}_{\text{goal}} \right\|_2^2.
\end{align*}
Since $\mathbf{A}(\mathbf{X})$ depends on the final state, we use an iterative procedure: at each step, we (1) fix $\mathbf{A}$ based on the current $\mathbf{X}$, (2) compute the control $\mathbf{C}^*$ via least squares, and (3) update the state using the resulting pressure. This converges in a few iterations in practice.

In our example, we optimize three internal pressure values ($k=3$) and enforce that the final position of a specific vertex matches a user-defined goal. Since the reduced state vector includes both deformation and momentum, we also require the final momentum to be nearly zero, consistent with the quasi-static assumption. To avoid non-physical negative pressures, we solve for $\mathbf{C}$ using non-negative least squares (NNLS); however, unconstrained least squares may be more appropriate in applications where bidirectional actuation is allowed. We perform this optimization iteratively: in each iteration, $\mathbf{A}(\mathbf{X})$ is updated based on the current deformation, and a new $\mathbf{C}$ is computed. We fix the number of iterations to 5.

By bypassing step-by-step simulation and directly optimizing in the reduced space, our method enables efficient long-horizon control with real-time performance.

As shown in Fig.~\ref{fig:finger_interactive}, the user specifies a target trajectory by drawing in a 2D interface. 
Given this target motion, the system solves for the pressure signal that drives the simulated finger to follow the desired trajectory. 
Because time stepping reduces to exponentiating a low-dimensional Koopman operator, this control optimization can be performed interactively as the target is edited.

To evaluate accuracy, we apply the same pressure signal to a full-space simulation and compare the resulting deformation to the prediction of our reduced model. 
Fig.~\ref{fig:finger_error} shows both the predicted and ground-truth trajectories, along with the per-frame reconstruction error. 
Despite the highly nonlinear deformation induced by pneumatic actuation, the error remains below 2\% for most frames and below 6\% for all frames, demonstrating that our method produces motion that closely matches the full simulation while enabling fast control.

Beyond interactive demonstration, this capability is directly relevant to robotics and design workflows, where fast forward and inverse simulation are essential for control, optimization, and rapid prototyping. 
By enabling efficient long-horizon prediction and real-time force estimation, our formulation provides a practical tool for exploring and controlling deformable mechanisms such as soft robots and pneumatic actuators.

\subsection{Fast Shape Optimization}

Our formulation enables efficient shape optimization by differentiating through long-horizon deformable dynamics. We demonstrate this on a pneumatically actuated frog model, where the geometry of the internal air chamber is parameterized by a 4-dimensional shape code (Figure~\ref{fig:frog_shape_gen}). The objective is to maximize upward deformation—specifically, the mean $y$-coordinate displacement of a target body region—after actuation by internal air pressure.

Using our Koopman-based reduced-order model, we perform shape optimization directly in the reduced space. Given a shape code $\gamma \in \mathbb{R}^4$, we compute the final state $\mathbf{X}_{\text{final}}(\gamma)$ using the iterative control method described in Section~\ref{sec:iteractive_control}, leveraging Equation~\ref{eq:air_final}. The loss is defined as the negative mean of the $y$-component of the deformation in the designated region. Gradients of this loss with respect to the shape code are approximated using finite differences with a small perturbation $\varepsilon=10^{-4}$.

We perform gradient descent with a learning rate of 0.025 to update the shape code. Because time integration is performed via exponentiation of a low-dimensional linear operator, this optimization remains efficient even over long horizons—unlike full-space simulation, where each gradient evaluation would require expensive recomputation of the full trajectory.

Figure~\ref{fig:frog_shape_opt} illustrates the shape optimization process. We initialize the frog with a default chamber geometry and optimize to maximize vertical lift. Over 150 iterations, the shape evolves to produce significantly larger deformations. When tested in a full-space simulator, the optimized geometry retains similar behavior, validating the reduced-space optimization.

\subsection{Learning Dynamics from Real-World Videos}

Another strength of our neural DMD framework is its ability to learn dynamics directly from real-world data.

We demonstrate this on two physical objects with distinct material properties: a soft, highly damped blob and a more elastic rubber duck. For each, we record two short RGB videos of the object responding to external perturbations. We apply a dot marker pattern to the surface and track their 2D motion using OpenCV’s \texttt{calcOpticalFlowPyrLK} across frames. From these tracked 2D trajectories, we reconstruct sparse state sequences $\mathbf{X}$ representing system evolution over time.

These trajectories are then used to train our neural DMD model. Crucially, our model uses coordinate-based neural fields to represent all basis functions and operators continuously over space, allowing it to generalize across mesh resolutions. This lets us reconstruct dense high-resolution deformations of the object at test time—even though the training data was sparse and noisy.

\begin{figure}
\centering
\includegraphics[width = 1.0\linewidth]{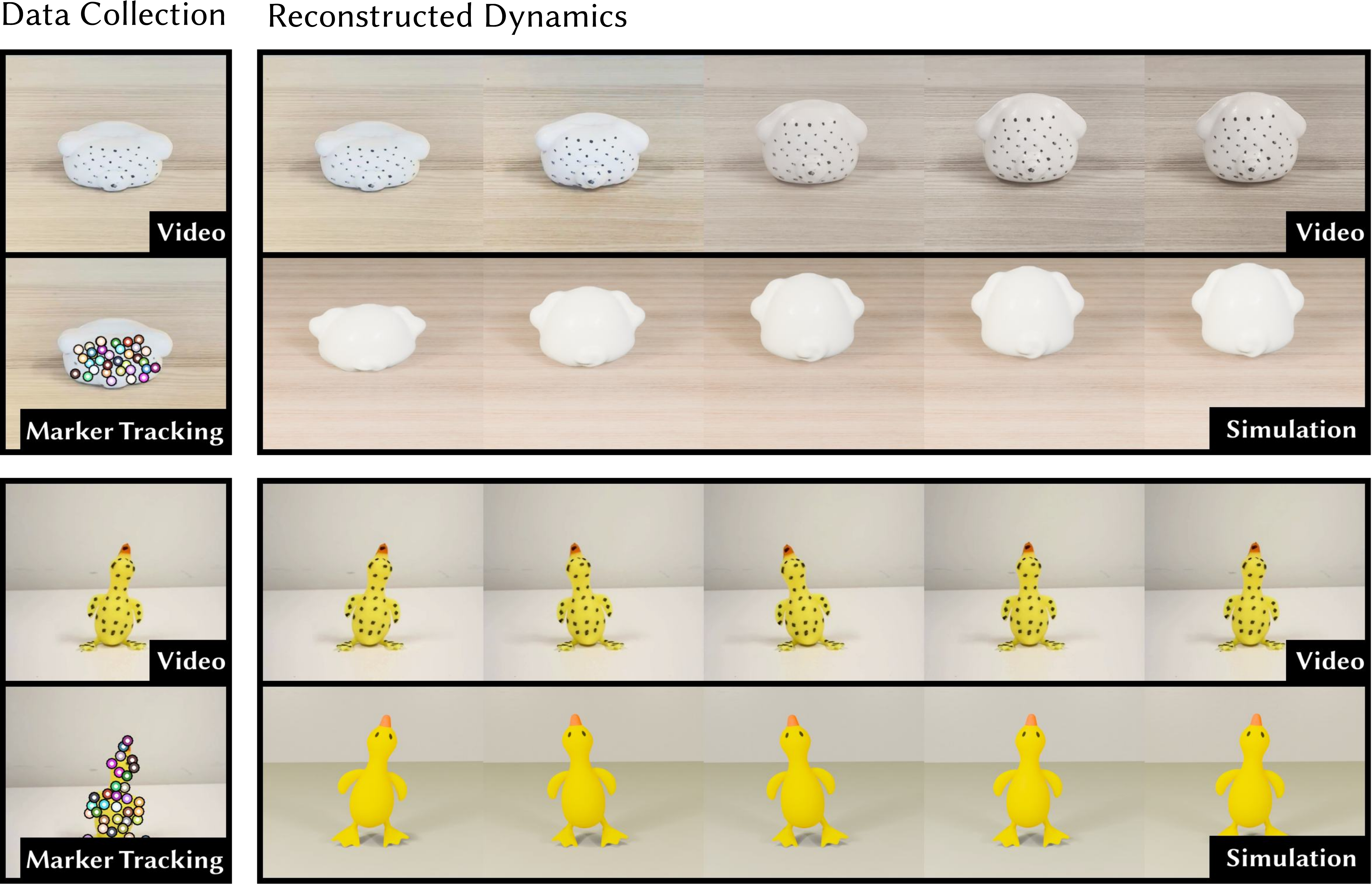}
\caption{\emph{Learning dynamics from real-world videos.}
We record videos of two deformable objects, a damped blob (top) and a more elastic rubber duck (bottom), responding to external poking. Using optical flow tracking, we extract sparse surface trajectories from dot markers to form training sequences. Our neural DMD model learns from these purely visual inputs without requiring simulations, and reconstructs high-resolution deformations that reflect the physical properties of each object.}
\label{fig:real_world}
\end{figure}

Figure~\ref{fig:real_world} shows a qualitative comparison between the captured video, tracked markers, and the reconstructed dynamics using our learned model. We get plausible deformations consistent with the material properties, such as higher damping in the blob and larger oscillations in the duck.

\section{Discussion and Future Work}

This work introduces a Koopman-based formulation for deformable simulation in computer graphics by adapting dynamic mode decomposition (DMD) to elastic dynamics. 
By representing time evolution as a low-rank linear operator, our method enables log-linear time stepping via matrix exponentiation, allowing fast long-horizon simulation, interactive control, and efficient optimization. 
The discretization-agnostic neural extension further allows a single model to operate across meshes of different resolutions and shapes, enabling applications such as real-time interaction, control-force estimation, and shape optimization that are difficult to achieve with purely numerical reduced models.

A key conceptual contribution of this work is to treat deformable dynamics not only as a spatially reduced system, but also as a temporally reduced one. 
Unlike conventional reduced-order models, which still rely on step-by-step numerical integration, our formulation directly advances the system through time using a learned linear operator. 
This separation of time stepping from spatial discretization opens new possibilities for fast simulation and inverse problems in deformable systems.

At the same time, our approach is fundamentally data-driven. 
The quality and stability of the learned Koopman operator depend on the coverage and fidelity of the training data, and the model cannot reliably extrapolate beyond the range of dynamics it has observed. 
In contrast, classical physics-based reduced-order models allow changes in material parameters, boundary conditions, or integrators to be applied analytically at runtime. 
Our learned model sacrifices some of this flexibility in exchange for fast time stepping and generalization across discretizations and shapes.


The dimensionality of the reduced space plays a key role in the performance of our method, but more dimensions do not necessarily lead to better results. \revision{For numerical models, Fig.~\ref{fig:whale_error} shows that increasing the number of basis vectors improves accuracy initially, but does not yield monotonic gains and can even slightly degrade performance at higher ranks. This behavior stems from the use of truncated SVD in DMD, which is optimal for reconstructing the data but not for estimating the inverse required by the operator; including modes associated with small singular values introduces ill-conditioning and amplifies errors. This issue is inherent to standard DMD and could potentially be mitigated by more robust variants (e.g., total least-squares or optimized DMD).} For neural models, a larger reduced space can make training harder to converge. Similar challenges with increased reduced-space dimensionality have also been observed in other neural ROMs~\cite{Sharp:2023:datafree}.

\revision{While the model generalizes well across moderate force variations, we observe mild degradation in more strongly nonlinear deformations, particularly when the training trajectories involve significant rotations or complex deformations. For instance, in the $1.33\times$ force case in Fig.~\ref{fig:armadillo_force_magnitude}, the armadillo exhibits slight scaling discrepancies in the body region. Similarly, }
\par\noindent
\begin{minipage}[t]{0.3\columnwidth}
\revision{under nonlinear rotational loading (inset), noticeable differences arise when the test rotation is twice that seen during training. In these regimes, extrapolation does }
\end{minipage}\hfill
\begin{minipage}[t]{0.65\columnwidth}
\vspace{-5pt}
\includegraphics[width=\columnwidth]{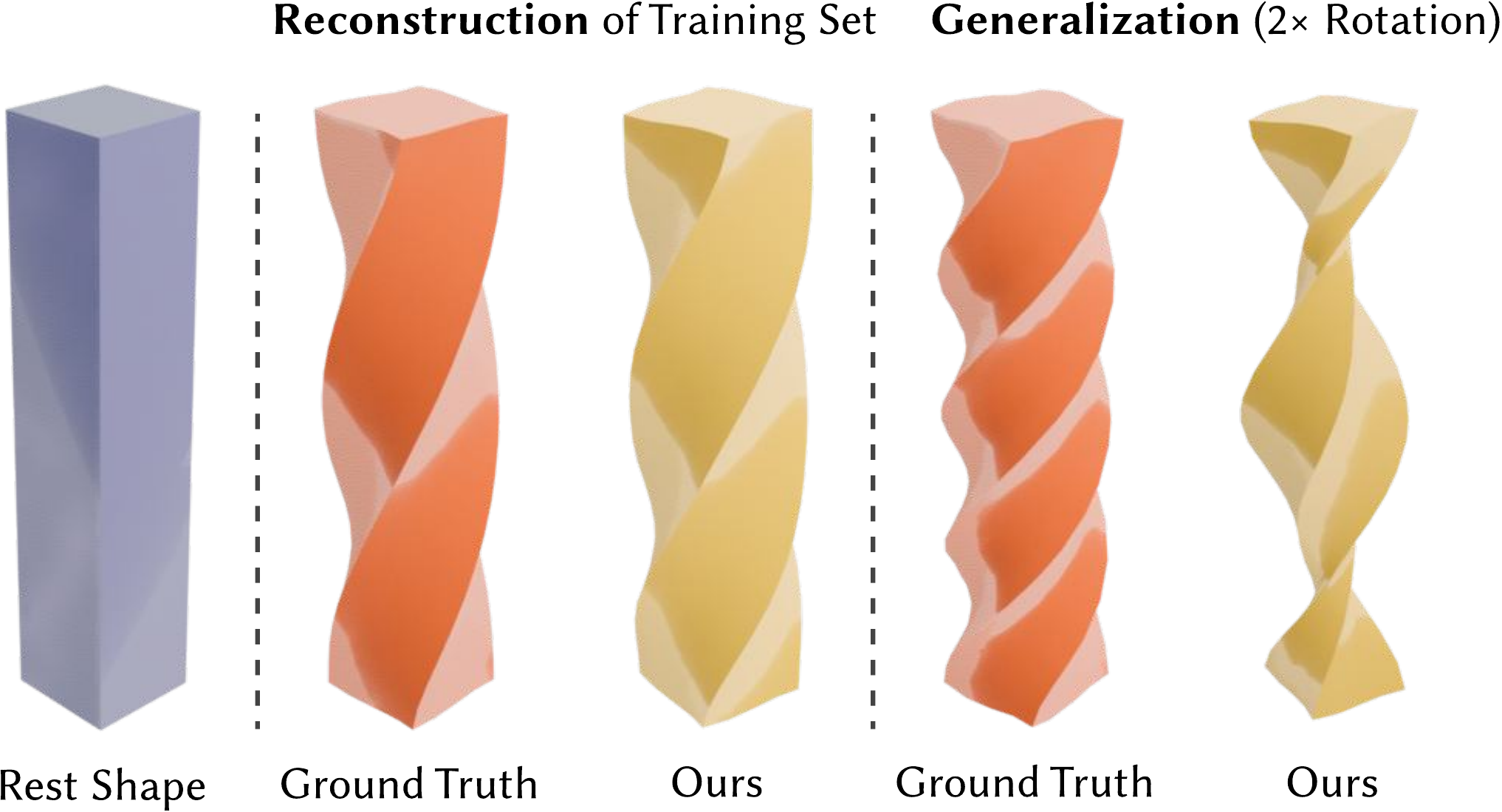}%
\end{minipage}
\par\vspace{3pt}\noindent 
\revision{not fully preserve deformation magnitude, leading to deviations from the ground truth.}

Looking forward, we believe that combining Koopman-based time reduction with physics-informed or hybrid training objectives could further improve robustness and extrapolation. 
Integrating explicit physical constraints, such as energy or momentum conservation, into the learning process may help stabilize long-term predictions. 

\revision{We do not consider contact forces in this work. A straightforward extension would be to incorporate penalty-based contact forces as external inputs in Eq.~\ref{eq:force}. We believe that handling contact is an interesting direction for future work.}

Finally, extending this framework to multi-body systems, and real-world sensor data would broaden its applicability to robotics, design, and interactive simulation.

\begin{acks}
We would like to thank our lab system administrator, John Hancock, and our financial officer, Xuan Dam, for their invaluable administrative support. We thank Towa Shixun Huang for proofreading assistance. We appreciate the valuable feedback from our reviewers, as well as from attendees of our group meetings at DGP and meetings at Meta Reality Labs. The first author would like to thank her family for providing the toys shown in Fig.~\ref{fig:real_world}. We gratefully acknowledge financial support from Meta. Finally, we acknowledge the support of the Natural Sciences and Engineering Research Council of Canada (NSERC) grant RGPIN-2021-03733.
\end{acks}

\bibliographystyle{ACM-Reference-Format}
\bibliography{main}


\begin{thebibliography}{50}


\ifx \showCODEN    \undefined \def \showCODEN     #1{\unskip}     \fi
\ifx \showISBNx    \undefined \def \showISBNx     #1{\unskip}     \fi
\ifx \showISBNxiii \undefined \def \showISBNxiii  #1{\unskip}     \fi
\ifx \showISSN     \undefined \def \showISSN      #1{\unskip}     \fi
\ifx \showLCCN     \undefined \def \showLCCN      #1{\unskip}     \fi
\ifx \shownote     \undefined \def \shownote      #1{#1}          \fi
\ifx \showarticletitle \undefined \def \showarticletitle #1{#1}   \fi
\ifx \showURL      \undefined \def \showURL       {\relax}        \fi
\providecommand\bibfield[2]{#2}
\providecommand\bibinfo[2]{#2}
\providecommand\natexlab[1]{#1}
\providecommand\showeprint[2][]{arXiv:#2}

\bibitem[An et~al\mbox{.}(2008)]%
        {An:Cubature:2008}
\bibfield{author}{\bibinfo{person}{Steven~S. An}, \bibinfo{person}{Theodore Kim}, {and} \bibinfo{person}{Doug~L. James}.} \bibinfo{year}{2008}\natexlab{}.
\newblock \showarticletitle{Optimizing cubature for efficient integration of subspace deformations}.
\newblock  \bibinfo{volume}{27}, \bibinfo{number}{5} (\bibinfo{year}{2008}).
\newblock
\showISSN{0730-0301}
\href{https://doi.org/10.1145/1409060.1409118}{doi:\nolinkurl{10.1145/1409060.1409118}}


\bibitem[Askham and Kutz(2018)]%
        {askham2018variable}
\bibfield{author}{\bibinfo{person}{Travis Askham} {and} \bibinfo{person}{J.~Nathan Kutz}.} \bibinfo{year}{2018}\natexlab{}.
\newblock \showarticletitle{Variable Projection Methods for an Optimized Dynamic Mode Decomposition}.
\newblock \bibinfo{journal}{\emph{SIAM Journal on Applied Dynamical Systems}} \bibinfo{volume}{17}, \bibinfo{number}{1} (\bibinfo{year}{2018}), \bibinfo{pages}{380--416}.
\newblock
\showeprint[arxiv]{1704.02343}~[math.NA]
\href{https://doi.org/10.1137/16M1124176}{doi:\nolinkurl{10.1137/16M1124176}}


\bibitem[Barbi{\v{c}} and James(2005)]%
        {barbivc2005real}
\bibfield{author}{\bibinfo{person}{Jernej Barbi{\v{c}}} {and} \bibinfo{person}{Doug~L James}.} \bibinfo{year}{2005}\natexlab{}.
\newblock \showarticletitle{Real-time subspace integration for St. Venant-Kirchhoff deformable models}.
\newblock \bibinfo{journal}{\emph{ACM transactions on graphics (TOG)}} \bibinfo{volume}{24}, \bibinfo{number}{3} (\bibinfo{year}{2005}), \bibinfo{pages}{982--990}.
\newblock


\bibitem[Barbi\u{c} and James(2010)]%
        {Barbic2010SubspaceSelfCollision}
\bibfield{author}{\bibinfo{person}{Jernej Barbi\u{c}} {and} \bibinfo{person}{Doug~L. James}.} \bibinfo{year}{2010}\natexlab{}.
\newblock \showarticletitle{Subspace Self-Collision Culling}.
\newblock \bibinfo{journal}{\emph{ACM Transactions on Graphics}} \bibinfo{volume}{29}, \bibinfo{number}{4} (\bibinfo{year}{2010}), \bibinfo{pages}{81:1--81:9}.
\newblock
\href{https://doi.org/10.1145/1778765.1778818}{doi:\nolinkurl{10.1145/1778765.1778818}}
\newblock
\shownote{Proc. SIGGRAPH}.


\bibitem[Benchekroun et~al\mbox{.}(2023)]%
        {benchekroun2023FastComplemDynamics}
\bibfield{author}{\bibinfo{person}{Otman Benchekroun}, \bibinfo{person}{Jiayi~Eris Zhang}, \bibinfo{person}{Siddhartha Chaudhuri}, \bibinfo{person}{Eitan Grinspun}, \bibinfo{person}{Yi Zhou}, {and} \bibinfo{person}{Alec Jacobson}.} \bibinfo{year}{2023}\natexlab{}.
\newblock \showarticletitle{Fast Complementary Dynamics via Skinning Eigenmodes}.
\newblock \bibinfo{journal}{\emph{ACM Transactions on Graphics}} (\bibinfo{year}{2023}).
\newblock


\bibitem[Bertiche et~al\mbox{.}(2022)]%
        {Bertiche:2022:NCS}
\bibfield{author}{\bibinfo{person}{Hugo Bertiche}, \bibinfo{person}{Meysam Madadi}, {and} \bibinfo{person}{Sergio Escalera}.} \bibinfo{year}{2022}\natexlab{}.
\newblock \showarticletitle{Neural Cloth Simulation}.
\newblock \bibinfo{journal}{\emph{ACM Trans. Graph.}} \bibinfo{volume}{41}, \bibinfo{number}{6}, Article \bibinfo{articleno}{220} (\bibinfo{date}{nov} \bibinfo{year}{2022}), \bibinfo{numpages}{14}~pages.
\newblock
\showISSN{0730-0301}
\href{https://doi.org/10.1145/3550454.3555491}{doi:\nolinkurl{10.1145/3550454.3555491}}


\bibitem[Cai et~al\mbox{.}(2022)]%
        {Cai:2022:CSDF}
\bibfield{author}{\bibinfo{person}{Xinhao Cai}, \bibinfo{person}{Eulalie Coevoet}, \bibinfo{person}{Alec Jacobson}, {and} \bibinfo{person}{Paul Kry}.} \bibinfo{year}{2022}\natexlab{}.
\newblock \showarticletitle{Active Learning Neural C-space Signed Distance Fields for Reduced Deformable Self-Collision}. In \bibinfo{booktitle}{\emph{Proceedings of Graphics Interface 2022}} (Montr{\'e}al, Quebec) \emph{(\bibinfo{series}{GI 2022})}. \bibinfo{publisher}{Canadian Information Processing Society}, \bibinfo{pages}{92 -- 100}.
\newblock
\showISSN{0713-5424}
\href{https://doi.org/10.20380/GI2022.11}{doi:\nolinkurl{10.20380/GI2022.11}}


\bibitem[Chang et~al\mbox{.}(2024)]%
        {chang2024neuralrepresentationshapedependentlaplacian}
\bibfield{author}{\bibinfo{person}{Yue Chang}, \bibinfo{person}{Otman Benchekroun}, \bibinfo{person}{Maurizio~M. Chiaramonte}, \bibinfo{person}{Peter~Yichen Chen}, {and} \bibinfo{person}{Eitan Grinspun}.} \bibinfo{year}{2024}\natexlab{}.
\newblock \bibinfo{title}{Shape Space Spectra}.
\newblock
\showeprint[arxiv]{2408.10099}~[cs.GR]
\urldef\tempurl%
\url{https://arxiv.org/abs/2408.10099}
\showURL{%
\tempurl}


\bibitem[Chang et~al\mbox{.}(2023)]%
        {chang:2023:licrom}
\bibfield{author}{\bibinfo{person}{Yue Chang}, \bibinfo{person}{Peter~Yichen Chen}, \bibinfo{person}{Zhecheng Wang}, \bibinfo{person}{Maurizio~M. Chiaramonte}, \bibinfo{person}{Kevin Carlberg}, {and} \bibinfo{person}{Eitan Grinspun}.} \bibinfo{year}{2023}\natexlab{}.
\newblock \showarticletitle{LiCROM: Linear-Subspace Continuous Reduced Order Modeling with Neural Fields}. In \bibinfo{booktitle}{\emph{SIGGRAPH Asia 2023 Conference Papers}} (, Sydney, NSW, Australia,) \emph{(\bibinfo{series}{SA '23})}. \bibinfo{publisher}{Association for Computing Machinery}, \bibinfo{address}{New York, NY, USA}, Article \bibinfo{articleno}{111}, \bibinfo{numpages}{12}~pages.
\newblock
\showISBNx{9798400703157}
\href{https://doi.org/10.1145/3610548.3618158}{doi:\nolinkurl{10.1145/3610548.3618158}}


\bibitem[Chen et~al\mbox{.}(2023a)]%
        {chenwu:2023:insr-pde}
\bibfield{author}{\bibinfo{person}{Honglin Chen}, \bibinfo{person}{Rundi Wu}, \bibinfo{person}{Eitan Grinspun}, \bibinfo{person}{Changxi Zheng}, {and} \bibinfo{person}{Peter~Yichen Chen}.} \bibinfo{year}{2023}\natexlab{a}.
\newblock \showarticletitle{Implicit Neural Spatial Representations for Time-dependent PDEs}. In \bibinfo{booktitle}{\emph{International Conference on Machine Learning}}.
\newblock


\bibitem[Chen et~al\mbox{.}(2023b)]%
        {chen2023crom}
\bibfield{author}{\bibinfo{person}{Peter~Yichen Chen}, \bibinfo{person}{Jinxu Xiang}, \bibinfo{person}{Dong~Heon Cho}, \bibinfo{person}{Yue Chang}, \bibinfo{person}{G~A Pershing}, \bibinfo{person}{Henrique~Teles Maia}, \bibinfo{person}{Maurizio~M Chiaramonte}, \bibinfo{person}{Kevin~Thomas Carlberg}, {and} \bibinfo{person}{Eitan Grinspun}.} \bibinfo{year}{2023}\natexlab{b}.
\newblock \showarticletitle{{CROM}: Continuous Reduced-Order Modeling of {PDE}s Using Implicit Neural Representations}. In \bibinfo{booktitle}{\emph{The Eleventh International Conference on Learning Representations}}.
\newblock
\urldef\tempurl%
\url{https://openreview.net/forum?id=FUORz1tG8Og}
\showURL{%
\tempurl}


\bibitem[Chen et~al\mbox{.}(2025)]%
        {chen2025dmd}
\bibfield{author}{\bibinfo{person}{Siyuan Chen}, \bibinfo{person}{Yixin Chen}, \bibinfo{person}{Jonathan Panuelos}, \bibinfo{person}{Otman Benchekroun}, \bibinfo{person}{Yue Chang}, \bibinfo{person}{Eitan Grinspun}, {and} \bibinfo{person}{Zhecheng Wang}.} \bibinfo{year}{2025}\natexlab{}.
\newblock \showarticletitle{Fast Subspace Fluid Simulation with a Temporally-Aware Basis}.
\newblock \bibinfo{journal}{\emph{ACM Transactions on Graphics}} (\bibinfo{year}{2025}).
\newblock


\bibitem[Chu et~al\mbox{.}(2022)]%
        {Chu2022Physics}
\bibfield{author}{\bibinfo{person}{Mengyu Chu}, \bibinfo{person}{Lingjie Liu}, \bibinfo{person}{Quan Zheng}, \bibinfo{person}{Erik Franz}, \bibinfo{person}{Hans-Peter Seidel}, \bibinfo{person}{Christian Theobalt}, {and} \bibinfo{person}{Rhaleb Zayer}.} \bibinfo{year}{2022}\natexlab{}.
\newblock \showarticletitle{Physics Informed Neural Fields for Smoke Reconstruction with Sparse Data}.
\newblock \bibinfo{journal}{\emph{ACM Transactions on Graphics, (Proc. SIGGRAPH)}} \bibinfo{volume}{41}, \bibinfo{number}{4} (\bibinfo{date}{aug} \bibinfo{year}{2022}), \bibinfo{pages}{119:1--119:15}.
\newblock


\bibitem[Cui et~al\mbox{.}(2018)]%
        {Cui2018Eigenfluids}
\bibfield{author}{\bibinfo{person}{Qiaodong Cui}, \bibinfo{person}{Pradeep Sen}, {and} \bibinfo{person}{Theodore Kim}.} \bibinfo{year}{2018}\natexlab{}.
\newblock \showarticletitle{Scalable Laplacian Eigenfluids}.
\newblock \bibinfo{journal}{\emph{ACM Transactions on Graphics}} \bibinfo{volume}{37}, \bibinfo{number}{4} (\bibinfo{year}{2018}).
\newblock
\newblock
\shownote{Proc. SIGGRAPH}.


\bibitem[de~Aguiar et~al\mbox{.}(2010)]%
        {deAguiar2010Stable}
\bibfield{author}{\bibinfo{person}{Edilson de Aguiar}, \bibinfo{person}{Leonid Sigal}, \bibinfo{person}{Adrien Treuille}, {and} \bibinfo{person}{Jessica~K. Hodgins}.} \bibinfo{year}{2010}\natexlab{}.
\newblock \showarticletitle{Stable Spaces for Real-Time Clothing}.
\newblock \bibinfo{journal}{\emph{ACM Trans. Graph. (SIGGRAPH)}} \bibinfo{volume}{29}, \bibinfo{number}{4}, Article \bibinfo{articleno}{106} (\bibinfo{year}{2010}).
\newblock
\href{https://doi.org/10.1145/1778765.1778843}{doi:\nolinkurl{10.1145/1778765.1778843}}


\bibitem[de~Witt et~al\mbox{.}(2012)]%
        {deWitt2012Eigenfunctions}
\bibfield{author}{\bibinfo{person}{Tyler de Witt}, \bibinfo{person}{Christian Lessig}, {and} \bibinfo{person}{Eugene Fiume}.} \bibinfo{year}{2012}\natexlab{}.
\newblock \showarticletitle{Fluid Simulation using Laplacian Eigenfunctions}.
\newblock \bibinfo{journal}{\emph{ACM Transactions on Graphics}} \bibinfo{volume}{31}, \bibinfo{number}{1} (\bibinfo{year}{2012}).
\newblock
\urldef\tempurl%
\url{https://www.dgp.toronto.edu/~tyler/fluids/}
\showURL{%
\tempurl}
\newblock
\shownote{Proc. SIGGRAPH}.


\bibitem[Deng et~al\mbox{.}(2023)]%
        {deng2023neural}
\bibfield{author}{\bibinfo{person}{Yitong Deng}, \bibinfo{person}{Hong-Xing Yu}, \bibinfo{person}{Diyang Zhang}, \bibinfo{person}{Jiajun Wu}, {and} \bibinfo{person}{Bo Zhu}.} \bibinfo{year}{2023}\natexlab{}.
\newblock \showarticletitle{Fluid Simulation on Neural Flow Maps}.
\newblock \bibinfo{journal}{\emph{ACM Trans. Graph.}} \bibinfo{volume}{42}, \bibinfo{number}{6} (\bibinfo{year}{2023}).
\newblock


\bibitem[Fulton et~al\mbox{.}(2019)]%
        {fulton2019latent}
\bibfield{author}{\bibinfo{person}{Lawson Fulton}, \bibinfo{person}{Vismay Modi}, \bibinfo{person}{David Duvenaud}, \bibinfo{person}{David~IW Levin}, {and} \bibinfo{person}{Alec Jacobson}.} \bibinfo{year}{2019}\natexlab{}.
\newblock \showarticletitle{Latent-space Dynamics for Reduced Deformable Simulation}. In \bibinfo{booktitle}{\emph{Computer graphics forum}}, Vol.~\bibinfo{volume}{38}. Wiley Online Library, \bibinfo{pages}{379--391}.
\newblock


\bibitem[Holden et~al\mbox{.}(2019)]%
        {Holden:2019:SNP}
\bibfield{author}{\bibinfo{person}{Daniel Holden}, \bibinfo{person}{Bang~Chi Duong}, \bibinfo{person}{Sayantan Datta}, {and} \bibinfo{person}{Derek Nowrouzezahrai}.} \bibinfo{year}{2019}\natexlab{}.
\newblock \showarticletitle{Subspace neural physics: fast data-driven interactive simulation}. In \bibinfo{booktitle}{\emph{Proceedings of the 18th Annual ACM SIGGRAPH/Eurographics Symposium on Computer Animation}} (Los Angeles, California) \emph{(\bibinfo{series}{SCA '19})}. \bibinfo{publisher}{Association for Computing Machinery}, \bibinfo{address}{New York, NY, USA}, Article \bibinfo{articleno}{6}, \bibinfo{numpages}{12}~pages.
\newblock
\showISBNx{9781450366779}
\href{https://doi.org/10.1145/3309486.3340245}{doi:\nolinkurl{10.1145/3309486.3340245}}


\bibitem[Ichinaga et~al\mbox{.}(2024)]%
        {Ichinaga2024}
\bibfield{author}{\bibinfo{person}{Sara~M. Ichinaga}, \bibinfo{person}{Francesco Andreuzzi}, \bibinfo{person}{Nicola Demo}, \bibinfo{person}{Marco Tezzele}, \bibinfo{person}{Karl Lapo}, \bibinfo{person}{Gianluigi Rozza}, \bibinfo{person}{Steven~L. Brunton}, {and} \bibinfo{person}{J.~Nathan Kutz}.} \bibinfo{year}{2024}\natexlab{}.
\newblock \showarticletitle{PyDMD: A Python Package for Robust Dynamic Mode Decomposition}.
\newblock \bibinfo{journal}{\emph{Journal of Machine Learning Research}} \bibinfo{volume}{25}, \bibinfo{number}{417} (\bibinfo{year}{2024}), \bibinfo{pages}{1--9}.
\newblock
\urldef\tempurl%
\url{http://jmlr.org/papers/v25/24-0739.html}
\showURL{%
\tempurl}


\bibitem[Jain et~al\mbox{.}(2024)]%
        {10.1145/3641519.3657438}
\bibfield{author}{\bibinfo{person}{Pranav Jain}, \bibinfo{person}{Ziyin Qu}, \bibinfo{person}{Peter~Yichen Chen}, {and} \bibinfo{person}{Oded Stein}.} \bibinfo{year}{2024}\natexlab{}.
\newblock \showarticletitle{Neural Monte Carlo Fluid Simulation}. In \bibinfo{booktitle}{\emph{ACM SIGGRAPH 2024 Conference Papers}} (Denver, CO, USA) \emph{(\bibinfo{series}{SIGGRAPH '24})}. \bibinfo{publisher}{Association for Computing Machinery}, \bibinfo{address}{New York, NY, USA}, Article \bibinfo{articleno}{9}, \bibinfo{numpages}{11}~pages.
\newblock
\showISBNx{9798400705250}
\href{https://doi.org/10.1145/3641519.3657438}{doi:\nolinkurl{10.1145/3641519.3657438}}


\bibitem[James et~al\mbox{.}(2006)]%
        {james2006precomputed}
\bibfield{author}{\bibinfo{person}{Doug~L James}, \bibinfo{person}{Jernej Barbi{\v{c}}}, {and} \bibinfo{person}{Dinesh~K Pai}.} \bibinfo{year}{2006}\natexlab{}.
\newblock \showarticletitle{Precomputed acoustic transfer: output-sensitive, accurate sound generation for geometrically complex vibration sources}.
\newblock \bibinfo{journal}{\emph{ACM Transactions on Graphics (TOG)}} \bibinfo{volume}{25}, \bibinfo{number}{3} (\bibinfo{year}{2006}), \bibinfo{pages}{987--995}.
\newblock


\bibitem[Kairanda et~al\mbox{.}(2024)]%
        {kair2024neuralclothsim}
\bibfield{author}{\bibinfo{person}{Navami Kairanda}, \bibinfo{person}{Marc Habermann}, \bibinfo{person}{Christian Theobalt}, {and} \bibinfo{person}{Vladislav Golyanik}.} \bibinfo{year}{2024}\natexlab{}.
\newblock \showarticletitle{NeuralClothSim: Neural Deformation Fields Meet the Thin Shell Theory}. In \bibinfo{booktitle}{\emph{Neural Information Processing Systems (NeurIPS)}}.
\newblock


\bibitem[Kim et~al\mbox{.}(2019)]%
        {kim2019deep}
\bibfield{author}{\bibinfo{person}{Byungsoo Kim}, \bibinfo{person}{Vinicius~C Azevedo}, \bibinfo{person}{Nils Thuerey}, \bibinfo{person}{Theodore Kim}, \bibinfo{person}{Markus Gross}, {and} \bibinfo{person}{Barbara Solenthaler}.} \bibinfo{year}{2019}\natexlab{}.
\newblock \showarticletitle{Deep fluids: A generative network for parameterized fluid simulations}. In \bibinfo{booktitle}{\emph{Computer Graphics Forum}}, Vol.~\bibinfo{volume}{38}. Wiley Online Library, \bibinfo{pages}{59--70}.
\newblock


\bibitem[Kim and James(2009)]%
        {Kim:Skipping:2009}
\bibfield{author}{\bibinfo{person}{Theodore Kim} {and} \bibinfo{person}{Doug~L. James}.} \bibinfo{year}{2009}\natexlab{}.
\newblock \showarticletitle{Skipping steps in deformable simulation with online model reduction}. \bibinfo{publisher}{Association for Computing Machinery}, \bibinfo{address}{New York, NY, USA}.
\newblock
\showISBNx{9781605588582}
\href{https://doi.org/10.1145/1661412.1618469}{doi:\nolinkurl{10.1145/1661412.1618469}}


\bibitem[Kutz et~al\mbox{.}(2015)]%
        {Kutz2015mrDMD}
\bibfield{author}{\bibinfo{person}{J.~Nathan Kutz}, \bibinfo{person}{Xing Fu}, \bibinfo{person}{Steven~L. Brunton}, {and} \bibinfo{person}{N.~Benjamin Erichson}.} \bibinfo{year}{2015}\natexlab{}.
\newblock \showarticletitle{Multi-Resolution Dynamic Mode Decomposition for Foreground/Background Separation and Object Tracking}. In \bibinfo{booktitle}{\emph{Proceedings of the IEEE International Conference on Computer Vision Workshops (ICCVW)}}. \bibinfo{pages}{921--929}.
\newblock
\href{https://doi.org/10.1109/ICCVW.2015.122}{doi:\nolinkurl{10.1109/ICCVW.2015.122}}


\bibitem[Li et~al\mbox{.}(2025)]%
        {Li2025LatentDynamics}
\bibfield{author}{\bibinfo{person}{Yue Li}, \bibinfo{person}{Gene Wei-Chin Lin}, \bibinfo{person}{Egor Larionov}, \bibinfo{person}{Aljaz Bozic}, \bibinfo{person}{Doug Roble}, \bibinfo{person}{Ladislav Kavan}, \bibinfo{person}{Stelian Coros}, \bibinfo{person}{Bernhard Thomaszewski}, \bibinfo{person}{Tuur Stuyck}, {and} \bibinfo{person}{Hsiao-Yu Chen}.} \bibinfo{year}{2025}\natexlab{}.
\newblock \showarticletitle{Self-Supervised Learning of Latent Space Dynamics}.
\newblock \bibinfo{journal}{\emph{Proceedings of the ACM on Computer Graphics and Interactive Techniques}} \bibinfo{volume}{8}, \bibinfo{number}{4}, Article \bibinfo{articleno}{57} (\bibinfo{year}{2025}), \bibinfo{numpages}{18}~pages.
\newblock
\href{https://doi.org/10.1145/3747854}{doi:\nolinkurl{10.1145/3747854}}


\bibitem[Lyu et~al\mbox{.}(2024)]%
        {lyu2024accelerate}
\bibfield{author}{\bibinfo{person}{Aoran Lyu}, \bibinfo{person}{Shixian Zhao}, \bibinfo{person}{Chuhua Xian}, \bibinfo{person}{Zhihao Cen}, \bibinfo{person}{Hongmin Cai}, {and} \bibinfo{person}{Guoxin Fang}.} \bibinfo{year}{2024}\natexlab{}.
\newblock \showarticletitle{Accelerate Neural Subspace-Based Reduced-Order Solver of Deformable Simulation by Lipschitz Optimization}.
\newblock \bibinfo{journal}{\emph{ACM Transactions on Graphics (TOG)}} \bibinfo{volume}{43}, \bibinfo{number}{6} (\bibinfo{year}{2024}), \bibinfo{pages}{1--10}.
\newblock


\bibitem[Matyka and Ollila(2021)]%
        {matyka2021pressure}
\bibfield{author}{\bibinfo{person}{Maciej Matyka} {and} \bibinfo{person}{Mark Ollila}.} \bibinfo{year}{2021}\natexlab{}.
\newblock \showarticletitle{Pressure Model of Soft Body Simulation}. In \bibinfo{booktitle}{\emph{Proceedings of the Conference (add if known)}}.
\newblock


\bibitem[Mezic(2020)]%
        {mezic2020koopman}
\bibfield{author}{\bibinfo{person}{Igor Mezic}.} \bibinfo{year}{2020}\natexlab{}.
\newblock \showarticletitle{Koopman Operator, Geometry, and Learning}.
\newblock \bibinfo{journal}{\emph{arXiv preprint arXiv:2010.05377}} (\bibinfo{year}{2020}).
\newblock
\showeprint[arxiv]{2010.05377}~[math.DS]
\href{https://doi.org/10.48550/arXiv.2010.05377}{doi:\nolinkurl{10.48550/arXiv.2010.05377}}


\bibitem[Modi et~al\mbox{.}(2024)]%
        {Modi:2024:Simplicits}
\bibfield{author}{\bibinfo{person}{Vismay Modi}, \bibinfo{person}{Nicholas Sharp}, \bibinfo{person}{Or Perel}, \bibinfo{person}{Shinjiro Sueda}, {and} \bibinfo{person}{David I.~W. Levin}.} \bibinfo{year}{2024}\natexlab{}.
\newblock \showarticletitle{Simplicits: Mesh-Free, Geometry-Agnostic Elastic Simulation}.
\newblock \bibinfo{journal}{\emph{ACM Trans. Graph.}} \bibinfo{volume}{43}, \bibinfo{number}{4}, Article \bibinfo{articleno}{117} (\bibinfo{date}{jul} \bibinfo{year}{2024}), \bibinfo{numpages}{11}~pages.
\newblock
\showISSN{0730-0301}
\href{https://doi.org/10.1145/3658184}{doi:\nolinkurl{10.1145/3658184}}


\bibitem[Mukherjee et~al\mbox{.}(2016)]%
        {Mukherjee:2016:IDSR}
\bibfield{author}{\bibinfo{person}{R. Mukherjee}, \bibinfo{person}{X. Wu}, {and} \bibinfo{person}{H. Wang}.} \bibinfo{year}{2016}\natexlab{}.
\newblock \showarticletitle{Incremental Deformation Subspace Reconstruction}.
\newblock \bibinfo{journal}{\emph{Comput. Graph. Forum}} \bibinfo{volume}{35}, \bibinfo{number}{7} (\bibinfo{date}{Oct.} \bibinfo{year}{2016}), \bibinfo{pages}{169–178}.
\newblock
\showISSN{0167-7055}


\bibitem[Proctor et~al\mbox{.}(2016)]%
        {proctor2016dmdc}
\bibfield{author}{\bibinfo{person}{Joshua~L. Proctor}, \bibinfo{person}{Steven~L. Brunton}, {and} \bibinfo{person}{J.~Nathan Kutz}.} \bibinfo{year}{2016}\natexlab{}.
\newblock \showarticletitle{Dynamic Mode Decomposition with Control}.
\newblock \bibinfo{journal}{\emph{SIAM Journal on Applied Dynamical Systems}} \bibinfo{volume}{15}, \bibinfo{number}{1} (\bibinfo{year}{2016}), \bibinfo{pages}{142--161}.
\newblock
\href{https://doi.org/10.1137/15M1013857}{doi:\nolinkurl{10.1137/15M1013857}}


\bibitem[Romero et~al\mbox{.}(2023)]%
        {romero2023contactdescriptorlearning}
\bibfield{author}{\bibinfo{person}{Cristian Romero}, \bibinfo{person}{Dan Casas}, \bibinfo{person}{Maurizio~M. Chiaramonte}, {and} \bibinfo{person}{Miguel~A. Otaduy}.} \bibinfo{year}{2023}\natexlab{}.
\newblock \showarticletitle{Learning Contact Deformations with General Collider Descriptors}. In \bibinfo{booktitle}{\emph{SIGGRAPH Asia 2023 Conference Papers}}. \bibinfo{publisher}{Association for Computing Machinery}, Article \bibinfo{articleno}{77}.
\newblock


\bibitem[Romero et~al\mbox{.}(2021)]%
        {Romero:LCCHSD:2021}
\bibfield{author}{\bibinfo{person}{Cristian Romero}, \bibinfo{person}{Dan Casas}, \bibinfo{person}{Jes\'{u}s P\'{e}rez}, {and} \bibinfo{person}{Miguel Otaduy}.} \bibinfo{year}{2021}\natexlab{}.
\newblock \showarticletitle{Learning Contact Corrections for Handle-Based Subspace Dynamics}.
\newblock \bibinfo{journal}{\emph{ACM Transactions on Graphics (TOG)}} \bibinfo{volume}{40}, \bibinfo{number}{4}, Article \bibinfo{articleno}{131} (\bibinfo{date}{jul} \bibinfo{year}{2021}), \bibinfo{numpages}{12}~pages.
\newblock
\showISSN{0730-0301}
\href{https://doi.org/10.1145/3450626.3459875}{doi:\nolinkurl{10.1145/3450626.3459875}}


\bibitem[Sashidhar and Kutz(2022)]%
        {sashidhar2022bagging}
\bibfield{author}{\bibinfo{person}{Diya Sashidhar} {and} \bibinfo{person}{J~Nathan Kutz}.} \bibinfo{year}{2022}\natexlab{}.
\newblock \showarticletitle{Bagging, optimized dynamic mode decomposition for robust, stable forecasting with spatial and temporal uncertainty quantification}.
\newblock \bibinfo{journal}{\emph{Philosophical Transactions of the Royal Society A}} \bibinfo{volume}{380}, \bibinfo{number}{2229} (\bibinfo{year}{2022}), \bibinfo{pages}{20210199}.
\newblock


\bibitem[Schmid(2010a)]%
        {schmid2010dmd}
\bibfield{author}{\bibinfo{person}{Peter~J. Schmid}.} \bibinfo{year}{2010}\natexlab{a}.
\newblock \showarticletitle{Dynamic mode decomposition of numerical and experimental data}.
\newblock \bibinfo{journal}{\emph{Journal of Fluid Mechanics}}  \bibinfo{volume}{656} (\bibinfo{year}{2010}), \bibinfo{pages}{5–28}.
\newblock
\href{https://doi.org/10.1017/S0022112010001217}{doi:\nolinkurl{10.1017/S0022112010001217}}


\bibitem[Schmid(2010b)]%
        {schmid2010dynamic}
\bibfield{author}{\bibinfo{person}{Peter~J Schmid}.} \bibinfo{year}{2010}\natexlab{b}.
\newblock \showarticletitle{Dynamic mode decomposition of numerical and experimental data}.
\newblock \bibinfo{journal}{\emph{Journal of fluid mechanics}}  \bibinfo{volume}{656} (\bibinfo{year}{2010}), \bibinfo{pages}{5--28}.
\newblock


\bibitem[Sharp et~al\mbox{.}(2023)]%
        {Sharp:2023:datafree}
\bibfield{author}{\bibinfo{person}{Nicholas Sharp}, \bibinfo{person}{Cristian Romero}, \bibinfo{person}{Alec Jacobson}, \bibinfo{person}{Etienne Vouga}, \bibinfo{person}{Paul Kry}, \bibinfo{person}{David~I.W. Levin}, {and} \bibinfo{person}{Justin Solomon}.} \bibinfo{year}{2023}\natexlab{}.
\newblock \showarticletitle{Data-Free Learning of Reduced-Order Kinematics}. In \bibinfo{booktitle}{\emph{ACM SIGGRAPH 2023 Conference Proceedings}} (Los Angeles, CA, USA) \emph{(\bibinfo{series}{SIGGRAPH '23})}. \bibinfo{publisher}{Association for Computing Machinery}, \bibinfo{address}{New York, NY, USA}, Article \bibinfo{articleno}{40}, \bibinfo{numpages}{9}~pages.
\newblock
\showISBNx{9798400701597}
\href{https://doi.org/10.1145/3588432.3591521}{doi:\nolinkurl{10.1145/3588432.3591521}}


\bibitem[Shen et~al\mbox{.}(2021)]%
        {shen2021high}
\bibfield{author}{\bibinfo{person}{Siyuan Shen}, \bibinfo{person}{Yang Yin}, \bibinfo{person}{Tianjia Shao}, \bibinfo{person}{He Wang}, \bibinfo{person}{Chenfanfu Jiang}, \bibinfo{person}{Lei Lan}, {and} \bibinfo{person}{Kun Zhou}.} \bibinfo{year}{2021}\natexlab{}.
\newblock \showarticletitle{High-order differentiable autoencoder for nonlinear model reduction}.
\newblock \bibinfo{journal}{\emph{arXiv preprint arXiv:2102.11026}} (\bibinfo{year}{2021}).
\newblock


\bibitem[Smith et~al\mbox{.}(2018)]%
        {Smith:2018:stablenh}
\bibfield{author}{\bibinfo{person}{Breannan Smith}, \bibinfo{person}{Fernando~De Goes}, {and} \bibinfo{person}{Theodore Kim}.} \bibinfo{year}{2018}\natexlab{}.
\newblock \showarticletitle{Stable Neo-Hookean Flesh Simulation}.
\newblock \bibinfo{journal}{\emph{ACM Trans. Graph.}} \bibinfo{volume}{37}, \bibinfo{number}{2}, Article \bibinfo{articleno}{12} (\bibinfo{date}{March} \bibinfo{year}{2018}), \bibinfo{numpages}{15}~pages.
\newblock
\showISSN{0730-0301}
\href{https://doi.org/10.1145/3180491}{doi:\nolinkurl{10.1145/3180491}}


\bibitem[Tao et~al\mbox{.}(2024)]%
        {tao2024neural}
\bibfield{author}{\bibinfo{person}{Yuanyuan Tao}, \bibinfo{person}{Ivan Puhachov}, \bibinfo{person}{Derek Nowrouzezahrai}, {and} \bibinfo{person}{Paul Kry}.} \bibinfo{year}{2024}\natexlab{}.
\newblock \showarticletitle{Neural Implicit Reduced Fluid Simulation}. In \bibinfo{booktitle}{\emph{SIGGRAPH Asia 2024 Conference Papers}}. \bibinfo{pages}{1--11}.
\newblock


\bibitem[Trusty et~al\mbox{.}(2023)]%
        {trusty:2023:subspacemfem}
\bibfield{author}{\bibinfo{person}{Ty Trusty}, \bibinfo{person}{Otman Benchekroun}, \bibinfo{person}{Eitan Grinspun}, \bibinfo{person}{Danny~M. Kaufman}, {and} \bibinfo{person}{David~I.W. Levin}.} \bibinfo{year}{2023}\natexlab{}.
\newblock \showarticletitle{Subspace Mixed Finite Elements for Real-Time Heterogeneous Elastodynamics}. In \bibinfo{booktitle}{\emph{SIGGRAPH Asia 2023 Conference Papers}} (Sydney, NSW, Australia) \emph{(\bibinfo{series}{SA '23})}. \bibinfo{publisher}{Association for Computing Machinery}, \bibinfo{address}{New York, NY, USA}, Article \bibinfo{articleno}{112}, \bibinfo{numpages}{10}~pages.
\newblock
\showISBNx{9798400703157}
\href{https://doi.org/10.1145/3610548.3618220}{doi:\nolinkurl{10.1145/3610548.3618220}}


\bibitem[von Tycowicz et~al\mbox{.}(2013)]%
        {Tycowicz:ECRDO:2013}
\bibfield{author}{\bibinfo{person}{Christoph von Tycowicz}, \bibinfo{person}{Christian Schulz}, \bibinfo{person}{Hans-Peter Seidel}, {and} \bibinfo{person}{Klaus Hildebrandt}.} \bibinfo{year}{2013}\natexlab{}.
\newblock \showarticletitle{An Efficient Construction of Reduced Deformable Objects}.
\newblock \bibinfo{journal}{\emph{ACM Transactions on Graphics (TOG)}} \bibinfo{volume}{32}, \bibinfo{number}{6}, Article \bibinfo{articleno}{213} (\bibinfo{date}{nov} \bibinfo{year}{2013}), \bibinfo{numpages}{10}~pages.
\newblock
\showISSN{0730-0301}
\href{https://doi.org/10.1145/2508363.2508392}{doi:\nolinkurl{10.1145/2508363.2508392}}


\bibitem[Wang et~al\mbox{.}(2024)]%
        {Wang2024PICT}
\bibfield{author}{\bibinfo{person}{Yiming Wang}, \bibinfo{person}{Siyu Tang}, {and} \bibinfo{person}{Mengyu Chu}.} \bibinfo{year}{2024}\natexlab{}.
\newblock \showarticletitle{Physics-Informed Learning of Characteristic Trajectories for Smoke Reconstruction}. In \bibinfo{booktitle}{\emph{ACM SIGGRAPH 2024 Conference Papers}} \emph{(\bibinfo{series}{SIGGRAPH '24})}. Article \bibinfo{articleno}{53}, \bibinfo{numpages}{11}~pages.
\newblock
\href{https://doi.org/10.1145/3641519.3657483}{doi:\nolinkurl{10.1145/3641519.3657483}}


\bibitem[Williams et~al\mbox{.}(2015)]%
        {williams2015edmd}
\bibfield{author}{\bibinfo{person}{Matthew~O. Williams}, \bibinfo{person}{Ioannis~G. Kevrekidis}, {and} \bibinfo{person}{Clarence~W. Rowley}.} \bibinfo{year}{2015}\natexlab{}.
\newblock \showarticletitle{A Data--Driven Approximation of the Koopman Operator: Extending Dynamic Mode Decomposition}.
\newblock \bibinfo{journal}{\emph{Journal of Nonlinear Science}} \bibinfo{volume}{25}, \bibinfo{number}{6} (\bibinfo{year}{2015}), \bibinfo{pages}{1307--1346}.
\newblock
\href{https://doi.org/10.1007/s00332-015-9258-5}{doi:\nolinkurl{10.1007/s00332-015-9258-5}}


\bibitem[Xu and Barbi\v{c}(2016)]%
        {Xu:2016:PSS}
\bibfield{author}{\bibinfo{person}{Hongyi Xu} {and} \bibinfo{person}{Jernej Barbi\v{c}}.} \bibinfo{year}{2016}\natexlab{}.
\newblock \showarticletitle{Pose-Space Subspace Dynamics}.
\newblock \bibinfo{journal}{\emph{ACM Trans. on Graphics (SIGGRAPH 2016)}} \bibinfo{volume}{35}, \bibinfo{number}{4} (\bibinfo{year}{2016}).
\newblock


\bibitem[Yang et~al\mbox{.}(2020)]%
        {yang2020learning}
\bibfield{author}{\bibinfo{person}{Shuqi Yang}, \bibinfo{person}{Xingzhe He}, {and} \bibinfo{person}{Bo Zhu}.} \bibinfo{year}{2020}\natexlab{}.
\newblock \showarticletitle{Learning Physical Constraints with Neural Projections}. In \bibinfo{booktitle}{\emph{Advances in Neural Information Processing Systems}}.
\newblock


\bibitem[Zesch et~al\mbox{.}(2023)]%
        {zesch2023neural}
\bibfield{author}{\bibinfo{person}{Ryan~S Zesch}, \bibinfo{person}{Vismay Modi}, \bibinfo{person}{Shinjiro Sueda}, {and} \bibinfo{person}{David~IW Levin}.} \bibinfo{year}{2023}\natexlab{}.
\newblock \showarticletitle{Neural Collision Fields for Triangle Primitives}. In \bibinfo{booktitle}{\emph{SIGGRAPH Asia 2023 Conference Papers}}. \bibinfo{pages}{1--10}.
\newblock


\bibitem[Zhang and Li(2024)]%
        {Zhang:2024:NGDSR}
\bibfield{author}{\bibinfo{person}{Meng Zhang} {and} \bibinfo{person}{Jun Li}.} \bibinfo{year}{2024}\natexlab{}.
\newblock \showarticletitle{Neural Garment Dynamic Super-Resolution}. In \bibinfo{booktitle}{\emph{SIGGRAPH Asia 2024 Conference Papers}} \emph{(\bibinfo{series}{SA ‘24})}. \bibinfo{publisher}{Association for Computing Machinery}, \bibinfo{address}{New York, NY, USA}, Article \bibinfo{articleno}{122}, \bibinfo{numpages}{11}~pages.
\newblock
\showISBNx{9798400711312}
\href{https://doi.org/10.1145/3680528.3687610}{doi:\nolinkurl{10.1145/3680528.3687610}}


\end{thebibliography}

\clearpage

\clearpage

\end{document}